\documentclass[reprint,NumberedRefs]{JASA}

\usepackage{siunitx}
\usepackage{filecontents}
\usepackage{rotating}
\usepackage{multirow}

\begin{document}
% PREPRINT: 
\title[Source identification in sparse beamforming maps]{PREPRINT: Automatic source localization and spectra generation from sparse beamforming maps}

% ie
%\title[JASA/Sample JASA Article]{Sample JASA Article}

%% repeat as needed
\author{A. Goudarzi}
\email{armin.goudarzi@dlr.de}
\author{C. Spehr}
\email{carsten.spehr@dlr.de}
\affiliation{Institute of Aerodynamics and Flow Technology, German Aerospace Center (DLR), Germany}

\author{S. Herbold}
\email{herbold@cs.uni-goettingen.de}
\affiliation{Institute of Computer Science, University of G\"ottingen, Germany}

% may be added after \author{}, ie
% \altaffiliation{Also at: Department1,  University1, City, State ZipCode, Country.}

%% for corresponding author

\begin{abstract}
This paper is part of a special issue on Machine Learning in Acoustics.\\
\\
Beamforming is an imaging tool for the investigation of aeroacoustic phenomena and results in high dimensional data that is broken down to spectra by integrating spatial Regions Of Interest. This paper presents two methods that enable the automated identification of aeroacoustic sources in sparse beamforming maps and the extraction of their corresponding spectra to overcome the manual definition of Regions Of Interest. The methods are evaluated on two scaled airframe half-model wind-tunnel measurements and on a generic monopole source. The first relies on the spatial normal distribution of aeroacoustic broadband sources in sparse beamforming maps. The second uses hierarchical clustering methods. Both methods are robust to statistical noise and predict the existence, location, and spatial probability estimation for sources based on which Regions Of Interest are automatically determined.
\end{abstract}

\maketitle

% comment in after preprint generation
% \section*{Nomenclature}
% \begin{tabular}{l l }
% 	$\alpha$ & Angle of attack in deg\\
% 	$f$ & Frequency in Hz\\
% 	He & Helmholtz number\\	
% 	$M$  & Mach number\\
% 	$s$  & Source-part\\
% 	$S$  & Source\\
% 	$\sigma$ & Standard deviation\\
% 	St & Strouhal number\\
% 	$T$ & Temperature in $\SI{}{\kelvin}$\\
% 	CLEAN-SC & CLEAN based on Source Coherence\\
% 	CSM & Cross-Spectral density Matrix\\
% 	DAMAS & Deconvolution Approach for the\\
% 	& Mapping of Acoustic Sources\\
% 	HDBSCAN & Hierarchical Density-Based Spatial\\
% 	& Clustering of Applications with Noise\\
% 	PDF & Probability Density Function\\
% 	PSD  & Power Spectrum Density in $\SI{}{\decibel\per\hertz}$ \\
% 	ROI & Region Of Interest\\
% 	SIND & Source Identification based on\\
% 	&spatial Normal Distribution\\ 
% 	SIHC & Source Identification based on\\
% 	&Hierarchical Clustering\\ 
% \end{tabular}
% \addtocounter{table}{-1}

\section{Introduction}
Multiple noise-generating phenomena and mechanisms exist in aeroacoustics~\citep{Mueller1979, Howe2007}. Expert domain knowledge and a detailed study of measurements are necessary to identify these phenomena in measurements. For the localization and investigation of aeroacoustic sources, microphone array beamforming is a reliable standard  method~\citep{Beamforming2019}. Beamforming measurements usually result in 2D or 3D beamforming maps for each observed frequency and are often varied over Mach number $M$, angle of attack of the flow $\alpha$, and geometrical parameters of the observed model. The level of the beamforming map entries indicates a sound source emission power, usually described by the Power Spectral Density (PSD$\left(\vec{x},f, M,\dots\right)$) \added{for each frequency $f$ and each focus point $\vec{x}$}, but can also result from background noise, spurious noise sources, and beamforming artifacts. Additionally, the localization can be disturbed by sound reflections, scattering, and refraction. Consequently, the resulting beamforming maps have to be analyzed to extract the desired source information. For this process, it is useful to integrate the high dimensional PSD\added[R2C18]{$(\vec{x},f)$} over spatial regions of the map to obtain \added{a} low-dimensional \deleted{data} PSD\added[R2C18]{$(f)$, that can be properly displayed in 2D}. Ideally, the process only includes the locations of the respective source of interest while rejecting locations of other sound sources. This is aggravated by the fact, that the source location may vary over the frequency and Mach number due to the flow-dependent nature of the sources themselves or due to the aforementioned scattering and refraction within the sound propagation from the source to the array microphones.\\

A common way to handle this source identification is the spatial integration of resulting beamforming maps over so-called Regions Of Interest (ROI). This results in low-dimensional data such as spectra~\citep{Martinez2019} which can be interpreted by human experts. There exist three approaches for the manual definition of ROIs. First, the whole beamforming map is integrated into a single spectrum which is then analyzed for prominent features, such as tones or peaks. Then, the beamforming map at these frequencies or frequency bands is observed to determine the origin of these sources, and ROIs are defined to account for these. Second, the beamforming maps are observed at a variety of chosen frequency intervals, and ROIs are defined based on the consistent appearance of sources at multiple frequencies, intuition, and experience. Third, ROIs are defined based on the studied geometry. A challenge for these methods is the distinction between beamforming artifacts and real sources; the correct separation of close and overlapping sources; the detection of sources with a low PSD and small-band sources; and the detection of sources that appear only at some of the measurement variations described above. The definition of the ROI may therefore not only depend on the wind-tunnel model but on the array resolution as well as the signal-to-noise ratio and the methods used to process the beamforming maps~\citep{Beamforming2019}. A wrong or insufficient ROI definition results in degraded or wrong spectra which is especially problematic since most of the following aeroacoustic analysis is based on these.\\

\added[G]{For aeroacoustic measurements,} the important task of defining ROIs is performed by the expert manually and takes typically from hours up to days from our experience, depending on the complexity of the beamforming maps and the studied model. For the identification of individual sources, machine learning proved to be a promising tool in acoustics, and Gaussian Mixture Models (GMM) were already deployed to track speaker sources in space-time~\citep{Bianco2019}. Another framework developed by Antoni~\citep{Antoni2012}, Zhang et al.~\citep{Zhang2012}, and Dong et al.~\citep{Dong2014,Dong2016} relied on the use of blind source separation (BSS). They showed that beamforming and nearfield acoustic holography can be reformulated as a BSS problem and specifically solved for incoherent sources. The authors provided a metric on how to determine the correct number of sources, which must be estimated before BSS, and showed that their method is robust towards an incorrect estimation. The BSS problem must be solved for each frequency and each measurement individually. However, as stated above, aeroacoustic sources often exist in limited frequency bands, at specific flow configurations, or can be only detected at specific angles of attack. While the authors suggest using spatial correlation analysis to identify which reconstructed source distribution belongs to which source at the corresponding frequency, the BSS approach itself lacks to provide a connection of the reconstructed sources over frequency. Thus, for a variety of measurement configuration, where the number of sources changes, BSS causes the same problem as beamforming, which is that the expert has to validate which reconstructed source distribution belongs to which source.\\

Even though these advanced techniques exist and conventional beamforming is in comparison very limited in terms of resolution and dynamic range it is still primarily used in wind-tunnel experiments in combination with deconvolution methods. The reasons for this are the low signal-to-noise ratio (SNR), which is often below $\langle\text{SNR}\rangle_f\le\SI{-10}{\decibel}$~\citep{Blacodon2011}, that it does not require prior knowledge of the source configurations and distributions, that it is robust and fast. \added[R1C02a, R1C03]{Thus, this paper focuses on methods that can be deployed after the use of various existing state-of-the-art imaging techniques in the frequency domain such as conventional beamforming in combination with CLEAN-SC or DAMAS.} The scaled air-frame models of a Dornier 728~\citep{Ahlefeldt2013} (Do728) and an Airbus A320~\citep{Ahlefeldt2017} are presented to derive these methods, discuss their usefulness, and specify a proof-of-concept implementation. The methods are then evaluated on a generic measurement, featuring three monopole sources with known location and source power.

\section{Datasets}
The data presented in this paper consists of beamforming measurements of two closed-section wind-tunnel models: one is of a Do728~\citep{Ahlefeldt2013} and one is of an A320~\citep{Ahlefeldt2017}; and a generic open-section wind-tunnel dataset which features a streamlined monopole (\SI{0.005}{\metre} opening) speaker as the primary noise source. For the Do728 dataset, values of  $\alpha_i=\SI{1}{\degree}$,$\SI{3}{\degree}$,$\SI{5}{\degree}$,$\SI{6}{\degree}$,$\SI{7}{\degree}$,$\SI{8}{\degree},\SI{9}{\degree},\SI{10}{\degree}$ are chosen for angle of attack $\alpha$ and $M_i=0.125$, $0.150$, $0.175$, $0.200$, $0.225$, $0.250$ for Mach number $M$. The mean Reynolds number is $\langle \text{Re} \rangle_M=\SI{1.4e6}{}$ based on the mean aerodynamic cord length $D_0=\SI{0.353}{\metre}$ and ambient temperature of $T=\SI{300}{\kelvin}$ at an ambient pressure $p_0=\SI{1e5}{\pascal}$. The array consists of 144 microphones at an aperture of $\SI{1.756}{\metre}\times\SI{1.3}{\metre}$ and \added{has} a sample frequency of $f_S=\SI{120}{\kilo\hertz}$. The A320 set contains $\alpha_i=\SI{3}{\degree}$,$\SI{7}{\degree}$,$\SI{7.15}{\degree}$,$\SI{9}{\degree}$, $M_i=0.175$, $0.200$, $0.225$ at a mean Reynolds number of $\langle \text{Re} \rangle_M=\SI{1.4e6}{}$ based on $D_0=\SI{0.308}{\metre}$, $T=\SI{300}{\kelvin}$, $p_0=\SI{1e5}{\pascal}$. The array consisted of 96 microphones at an aperture of $\SI{1.06}{\metre}\times\SI{0.5704}{\metre}$ and $f_S=\SI{150}{\kilo\hertz}$. The generic dataset consists of three individual speaker positions with unique, band-limited white noise. Mach numbers of $M_i=0.00$, $0.03$, and $0.06$ were chosen at ambient pressure $p_0=\SI{1e5}{\pascal}$ and temperature $T=\SI{300}{\kelvin}$ and for each flow configuration an additional noise-floor measurement was obtained, by turning off the speaker. The square, equidistant array consisted of $7\times7=49$ microphones with an aperture of $\SI{0.54}{\metre}\times\SI{0.54}{\metre}$ and was placed $\Delta z=\SI{0.65}{\metre}$ away from the sources. The sample rate was $f_S=\SI{32768}{\hertz}$. Thus, the Do728 dataset consists of 48 measurements, the A320 dataset consists of 12 measurements. For the generic measurement, the measurements of individual source positions are superpositioned and thus, the generic dataset consists of effectively three measurements (with the speaker turned on). Cross-Spectral density Matrices (CSM) are calculated using Welch's method with a block size of 1024 samples for the Do728, 512 samples for the A320, and 256 samples for the generic dataset, with $\SI{50}{\percent}$ overlap. The beamforming is performed using conventional beamforming and CLEAN-SC deconvolution with a focus point resolution of $\Delta x_1 = \Delta x_2 = \SI{5e-3}{\metre}$.

\section{Source identification}\label{sec:source_identification}

\begin{figure}
	\centering
	\includegraphics[width=\reprintcolumnwidth]{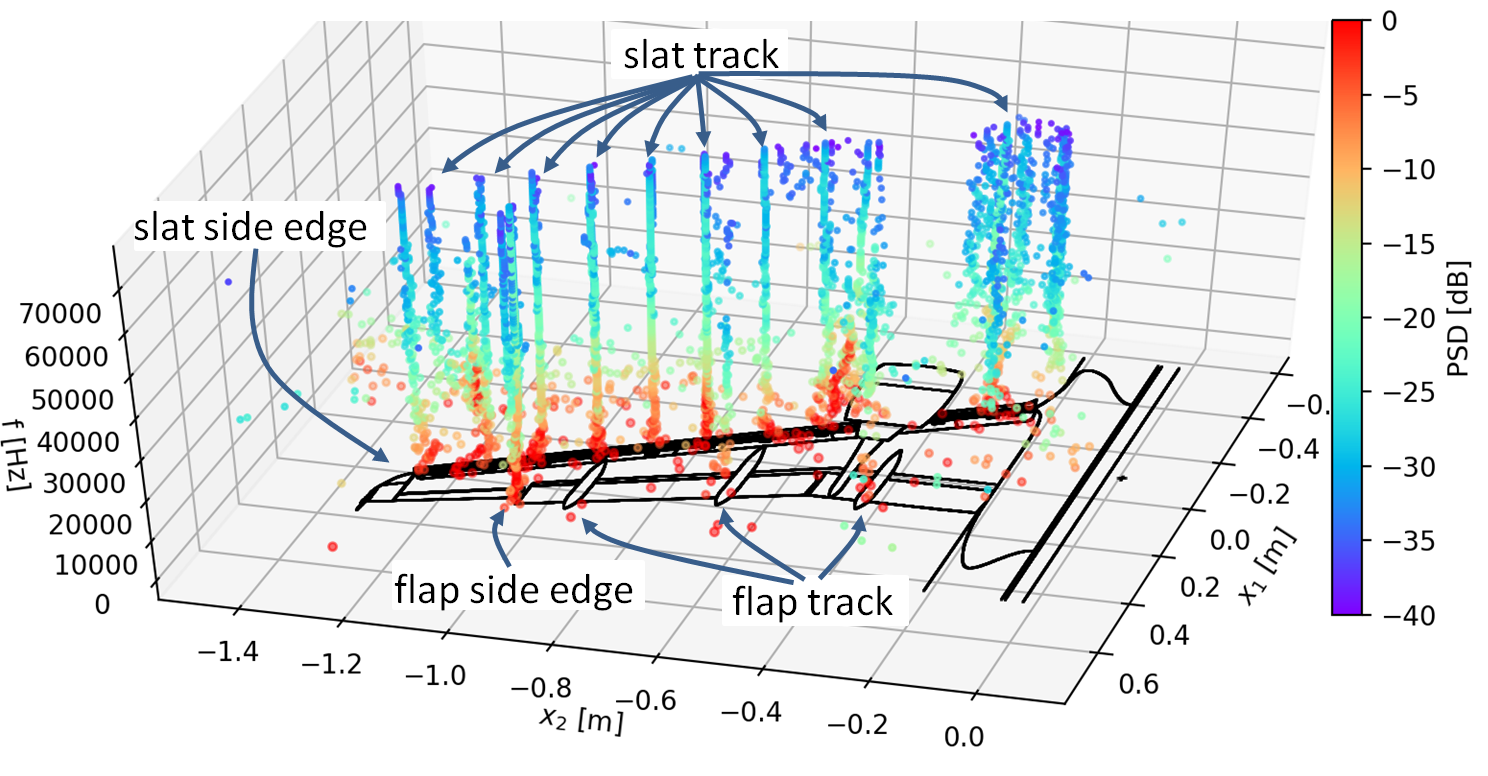}
	\caption{(Color online) A320, CLEAN-SC result on 2D-plane using conventional beamforming, the z-axis displays the frequency. The color represents the normalized PSD in decibel at $M=0.2$, $\alpha=\SI{3}{\degree}$.}
	\label{fig:Figure01}
\end{figure}

A general problem concerning beamforming is that at long wavelengths the localization of acoustic sources is difficult. Furthermore, imaging artifacts may occur due to the sparse spatial distribution of the microphone array. These artifacts result from background noise, the array's Point Spread Function, and aliasing or insufficient Welch estimations~\citep{Beamforming2019}.\\

In this part of the paper, we discuss two ideas on how to identify sources from beamforming maps contaminated with noise and obtain their spectrum. The ideas are based on the sparsity of beamforming maps, in the sense that the quantity of zero elements ($\text{PSD}=\SI{0}{\pascal\squared\per\hertz}$) is large compared to the non-zero elements in the maps ($\text{PSD}\ge\SI{0}{\pascal\squared\per\hertz}$). This can be achieved using inverse beamforming methods or conventional beamforming~\citep{Beamforming2019} in combination with what is known in the aeroacoustic beamforming community as ``deconvolution'', such as CLEAN-SC~\citep{Sijtsma2007} or DAMAS~\citep{Bahr2017}.\\

For this paper, we choose conventional beamforming with diagonal removal~\citep{Beamforming2019} in combination with CLEAN-SC over DAMAS, because of the huge number of computed beamforming maps and the high spatial resolution of the maps. CLEAN-SC assumes point-like sources and then subtracts coherent portions of the dirty beamforming map~\citep{Sijtsma2007}. This removes most of the Point Spread Function but will also result in a single $\text{PSD}(\vec{x}_0,f_0)$ representation of spatially distributed or correlated sources, that is only defined in a single spatial location $x_0$ for a given frequency $f_0$. We make this an advantage as this results in extremely sparse representations of the source map, which allows us to analyze the spatial distributions of non-zero elements in space and frequency. For terminology, we call every non-zero element in the map a source-part $s$, since once they are integrated over space and frequency they represent full sources. Thus, the resulting sparse beamforming maps can be reduced to a list of source-part vectors $s_i=[\vec{x}_i,f_i,\alpha_i,M_i,\text{PSD}_i]$.\\

Figure~\ref{fig:Figure01} displays the source-parts of the CLEAN-SC result on a 2D-focus grid for the A320. On the z-axis, the frequency is displayed, the color represents the normalized PSD. We can identify multiple vertical pillars of source-parts $s$ which, spatially integrated, represent a source spectrum $\text{PSD}(f)$. However, we also observe pillars that suddenly split with increasing frequency (e.g., at the flap side edge) or dense point clouds that are spatially scattered around (e.g., the inner slat). A source-part pillar that splits with increasing frequency can either be caused by a complex aeroacoustic mechanism or the limitations of beamforming and CLEAN-SC. It is expected to observe this behavior for frequencies that are around the Rayleigh Criterion $f_R$ below which two separate sources cannot be spatially resolved. This frequency is in the range of $\SI{5}{\kilo\hertz}\ge f_R\le \SI{6}{\kilo\hertz}$ for the Do728 and $\SI{8}{\kilo\hertz}\ge f_R\le \SI{16}{\kilo\hertz}$ based on the oval array apertures and the distance between the high frequency pillars. Since the frequencies at which the pillars separate coincide with the Rayleigh frequencies $f_R$, we assume this behavior is caused by the latter. Unfortunately, beamforming and deconvolution methods do not provide any information on which source-parts (in space and frequency) are generated by the same turbulence-induced aeroacoustic source-mechanism.\\

Thus, up to now, large spatial ROIs were defined manually as integration areas to obtain spectra~\citep{Sijtsma2004} such as the whole slat and flap region. This partly contradicts the beamforming idea, as we often do not know where sources are located and whether all source-parts within the integration region belong to the same source. In the following part, we introduce two methods on how to estimate the existence and positions of individual sources in sparse beamforming maps and how to correctly assign the corresponding source-parts to them.

\subsection{Source Identification based on spatial Normal Distributions (SIND)}\label{sec:source_ident_normal}

\begin{figure}[h]
	\centering
	\includegraphics[width=\reprintcolumnwidth]{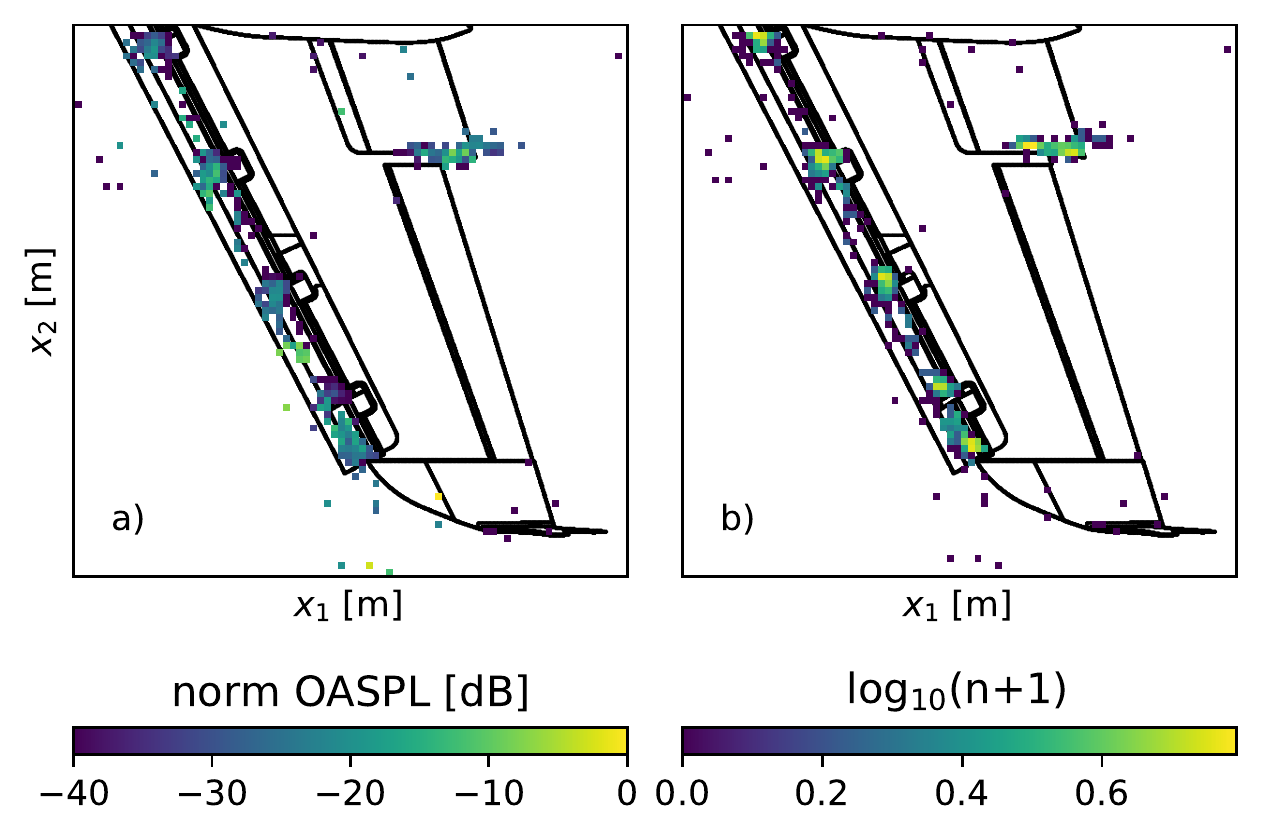}
	\caption{(Color online) A320, section of the CLEAN-SC map at $M=0.175, \alpha=\SI{3}{\degree}$. $a)$ shows the normalized OASPL, $b)$ shows a log-histogram of the $n$ source-parts $s$ per focus point $\vec{x}$.}
	\label{fig:Figure02}
\end{figure}

Figure~\ref{fig:Figure02} $a)$ shows the normalized Overall Sound Pressure Level (OASPL) for each individual spatial location $\vec{x}$ for the A320. The OASPL is the integration of the source-parts Sound Pressure Level (SPL) over frequency. We observe that the individual slat tracks, which we assume to be individual sources, cannot be easily distinguished based on the OASPL because the sound carries most energy at long wavelengths. Due to the array resolution, beamforming is not able to localize sources well at these wavelengths (see Figure~\ref{fig:Figure01}). However, ignoring the SPL and simply counting how often a source-part $s$ was reconstructed by CLEAN-SC at every location $\vec{x}$ in the entire map over frequency provides a better grasp on individual source distributions, which are shown in the logarithmic histogram in Figure~\ref{fig:Figure02} $b)$. Thus, the OASPL($\vec{x}$) gives an estimation of a source emission power while the histogram($\vec{x}$) gives an estimation of a frequency-interval or number of frequencies of the emission per location in the source map. In the log-histogram, we see mostly distinguishable blobs with maxima in their center that probably represent aeroacoustic sources, as the blobs' positions coincide with the location of the slat tracks, the slat side edge, and the flap side edge. Due to the Gaussian nature of the turbulence induced source mechanisms and the scattering and refraction of sound waves in turbulent structures~\citep{Ernst2020} we assume these blobs to be point-like sources that are smeared out in the beamforming map with maxima at locations that vary due to the mentioned phenomena.\\

While the blobs in the log-histogram do resemble normal distributions, statistical tests such as the Shapiro-Wilk or the Anderson-Darling test do not determine that data as normal. The reason for this is the discrete spatial sampling, the overlapping of sources, as well as the large population of source-parts. Thus, to verify the normality assumption, we compare the histogram of individual sources to a normal distribution. First, we fit a normal distribution to the log-distribution of the appearance of source-parts by minimizing the absolute difference between the source-part's position histogram and the estimated normal distribution using a L1-norm. Then, we compare the estimated distribution with the observed data. The normal distribution in 2D is calculated with eq.~\ref{eq:normal_dist}~\citep{Abramowitz1974}. For practical applications, we recommend optimizing for the normal distribution's amplitude $\hat{A}$, the standard deviations $\sigma_{x_i}$, the distribution rotation $\theta$, and the location $x_{i,0}$ by using a bounded optimization method with equations~\ref{eq:normal_dist_app}. The histogram's global maximum determines the starting values for the first source's $\hat{A},\vec{x}_{0}$; the bounds $\hat{A}\pm\varepsilon_{\hat{A}} ,\vec{x}\pm\varepsilon_{\vec{x}}$ prevent the optimizer from wandering off to a completely different source.

% \begin{equation}\label{eq:normal_dist}
\begin{multline}\label{eq:normal_dist}
N(x_1,x_2)=\hat{A}\exp\bigg(-\big(a(x_1-x_{1,0})^2\\+2b(x_1-x_{1,0})(x_2-x_{2,0})) + c(x_2-x_{2,0})^2\big)\bigg)
\end{multline}
% \end{equation}

\begin{subequations}\label{eq:normal_dist_app}
\begin{alignat}{3}
a&=& &\frac{\cos^2\theta}{2\sigma^2_{x_1}} &+& \frac{\sin^2\theta}{2\sigma^2_{x_2}}\\
b&=&-&\frac{\sin 2\theta}{4\sigma^2_{x_1}} &+& \frac{\sin 2\theta}{4\sigma^2_{x_2}}\\
c&=& &\frac{\sin^2\theta}{2\sigma^2_{x_1}} &+& \frac{\cos^2\theta}{2\sigma^2_{x_2}}
\end{alignat}
\end{subequations}

\begin{figure}
	\centering
	\includegraphics[width=\reprintcolumnwidth]{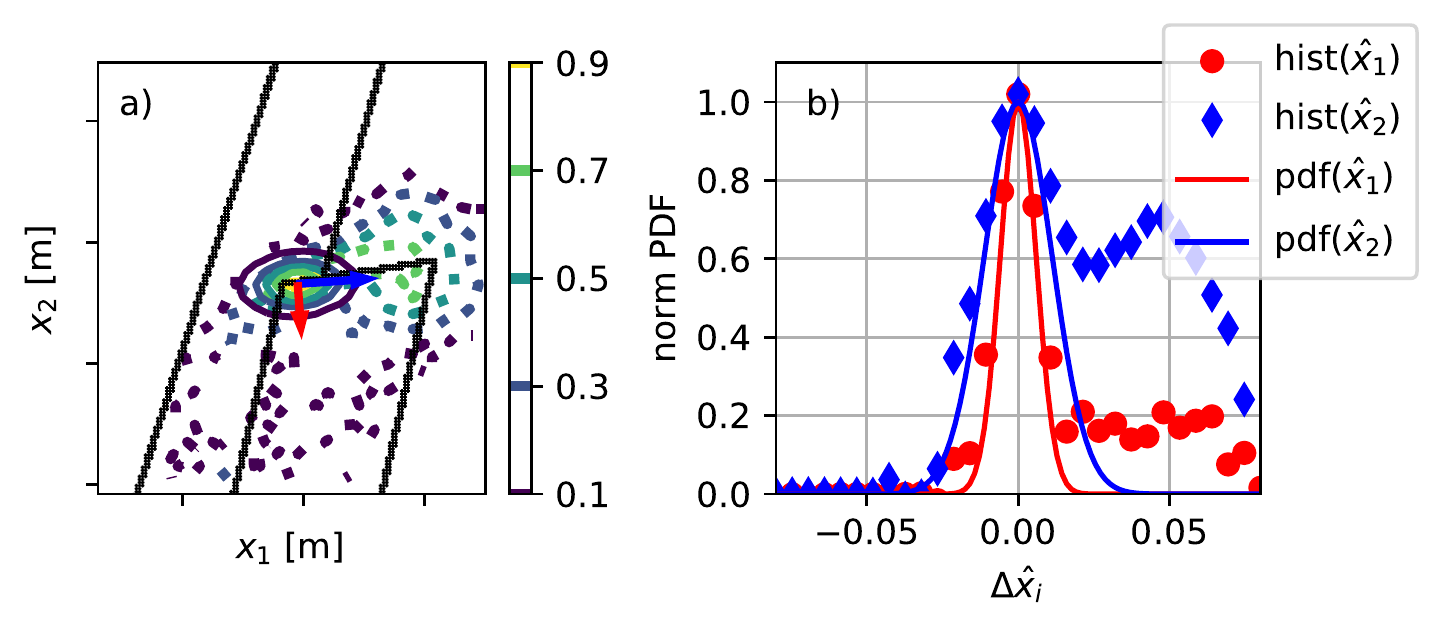}
	\caption{(Color online) Do728, flap side edge region. $a)$ shows the isocontour lines of the by $\hat{A}$ normalized distribution (dotted lines) and its fitted PDF (full lines). $b)$ shows the normalized distribution and PDF on its principal axis $\hat{x}_1$ and $\hat{x}_2$ which result from the $\theta$-rotation of the fitted distribution and are marked with arrows of the same color in $a)$.}
	\label{fig:Figure03}
\end{figure}

Figure~\ref{fig:Figure03} $a)$ shows the normalized log-distribution of the source-parts (dotted lines) for the Do728 flap side edge region. The histogram shows the summation of all source-part from all beamforming maps in the dataset, which contains 48 individual measurement configurations. We can determine two overlapping blobs in this region, a major one upstream and a minor one downstream. As described above,  a 2D normal distribution is fitted to minimize the major source-part blob (full lines) using eq.~\ref{eq:normal_dist}. We introduce two principal axes $\hat{x}_1$ and $\hat{x}_2$ for which the normal distributions standard deviations $\sigma_{x_i}$ are independent. They are obtained for each source from the fitted normal distribution's angle $\theta$. Figure~\ref{fig:Figure03} $b)$ shows the comparison of the normalized histogram and fitted distribution along these axes to verify the normality assumption. For terminology, we refer to the fitted, amplitude-normalized normal distributions as the Probability Density Function (PDF) of the source-part distribution of a source. While a PDF in the traditional mathematical sense is defined as a normalized distribution so that its integrated area is unity, our PDF is normalized so that $0\ge\text{PDF}(\vec{x})\ge1$. This means that the integrated area of our defined source-part PDF can be any real number $\mathbb{R}\ge0$ and that the PDF can be interpreted as the probability of a spatial location $\vec{x}$ belonging to a source.\\

As shown in Figure~\ref{fig:Figure03}, an individual source can be approximated with a normal distribution in the histogram. To find and fit all sources in the beamforming map (e.g., the second source on the right in Figure~\ref{fig:Figure03}), we introduce the distance metric $d_{S_i}$, see eq.~\ref{eq:dist_metric}, to measure and minimize the L1-norm of the estimated PDF of a source $S_i$ and the histogram. With the set $X_{S_i}$ containing all focus points $\vec{x}_j$, we want to minimize $d_{S_i}$ for all assumed sources $S_i\in S$ in the beamforming map so that the L1-norm of the source-part histogram and the fitted normal distributions achieves a minimum.

\begin{equation}\label{eq:dist_metric}
d_{S_i}=\sum_{\vec{x}_j \in X_{S_i}} \left|\text{hist}(\vec{x}_j)-\text{PDF}_{S_i}(\vec{x}_j)\right|
\end{equation}

Using this metric we can implement a greedy algorithm that finds all sources by minimizing $d_S=\sum_i d_{S_i}$ by iteration. First, we find the maximum in the source-part histogram; second, we fit a normal distribution that minimizes the histogram, see eq.~\ref{eq:dist_metric}; third, we subtract the fitted distribution from the histogram and forth, repeat the process until the remaining histogram-maximum drops below a threshold $t_I$. This threshold represents a lower significance bound and prevents endless fitting iterations since $d_S$ will decrease with an increasing number of sources that are either irrelevant or fitting artifacts. Thus, the order in which the method identifies sources in the histogram corresponds to their descending magnitude $\hat{A}_{S_i}$ in the histogram. Note, that this magnitude $\hat{A}_{S_i}$ does not explicitly depend on the source-part's PSD and thus, does not necessarily indicate a dominant source. Instead, a large $\hat{A}_{S_i}$ indicates either a broad-band source, a spatially well-localized source, or a combination of these features. However, since CLEAN-SC works within a certain SNR range, a set of source-parts that represent a source implicitly indicate, that the source was somewhat relevant within the beamforming map.\\

This makes this method similar to an iterative GMM. Traditionally, a degree-of-freedom weighted residual such as the Bayesian Information Criterion (BIC) is used for GMM, to determine the optimal number of sources~\citep{Schwarz1978}. Since the result of GMM heavily relies on the chosen number of sources, the number of sources must be estimated before clustering. However, this is not the case for this method, since it does not fit the source-part distributions (sources) simultaneously but works iteratively. Instead, the correct number of sources can be determined after the fitting process is complete. To do so, we integrate the fitted normal distributions, see eq.\ref{eq:normal_dist}, to obtain an area $A_{S_i}$ 
\begin{equation}
    A_{S_i}=\int_{x_1}\int_{x_2}\hat{A}\mathrm{PDF}_{S_i}(x_1,x_2)d x_2 d x_1
\end{equation}
for each source. This area reflects the impact of the estimated, individual sources on the L1-norm for $d_S$. If $A_{S_i}$ drops below a threshold $t_A$ we can reject it as a fitting artifact or negligible source. $A_{S_i}$ of artifacts is orders of magnitude below $A_{S_i}$ of real sources. However, if the threshold $t_I$ is set sufficiently high, SIND's iterations often stop before fitting artifacts occur.\\

\begin{figure}
	\centering
    \includegraphics[width=\reprintcolumnwidth]{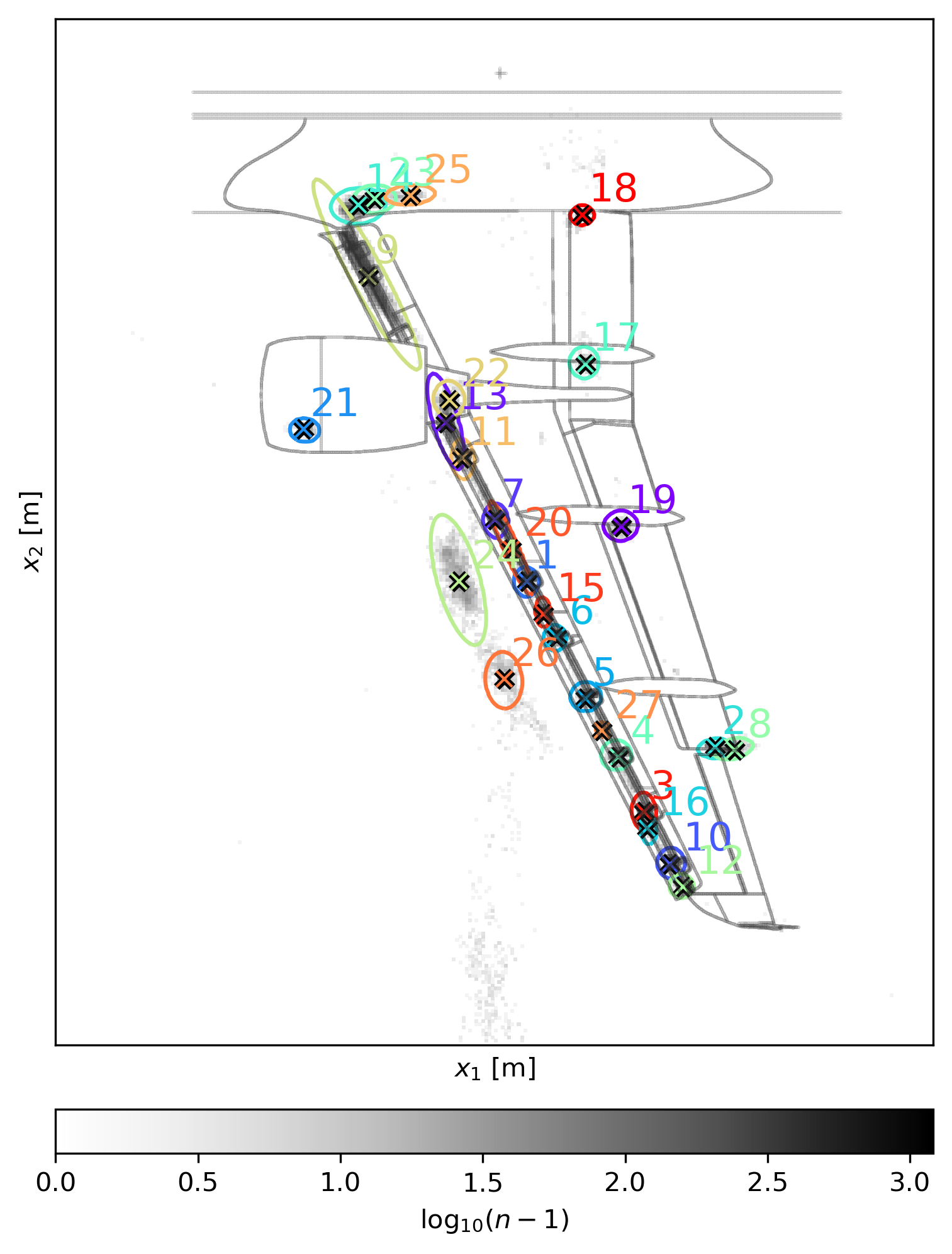}
	\caption{(Color online) A320. The SIND solution for $t_I=20$ is shown. The source numbers correspond to the order of found sources via the maxima in the histogram, which is displayed with the underlying colormap. The ellipses around the sources represent the PDF functions at $1-3\sigma$.}
	\label{fig:Figure04}
\end{figure}

\begin{figure}
	\centering
    \includegraphics[width=\reprintcolumnwidth]{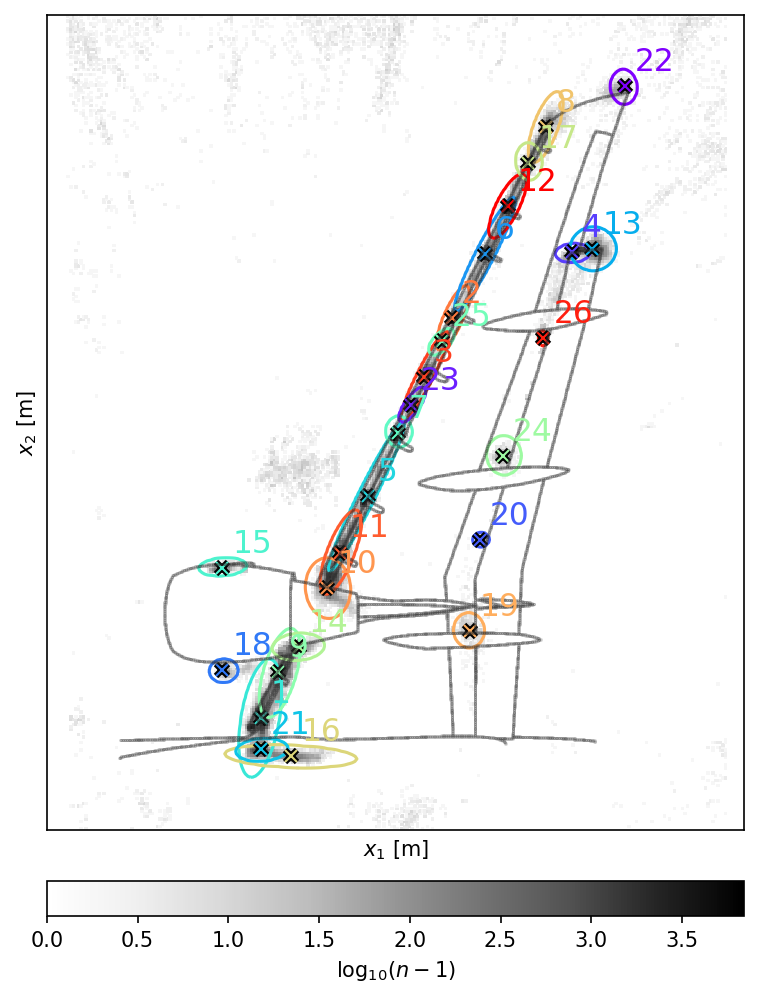}
	\caption{(Color online) Do728. The SIND solution for $t_I=30$ is shown. The source numbers correspond to the order of found sources via the maxima in the histogram, which is displayed with the underlying colormap. The ellipses around the sources represent the PDF functions at $1-3\sigma$.}
	\label{fig:Figure05}
\end{figure}

\begin{table}[ht]
\caption{\label{tab:Table01}SIND and SIHC parameters for the A320, Do728, and generic dataset and the total number $n$ of source-parts present in the datasets.}
\begin{ruledtabular}
\begin{tabular}{c|c|ccc|cc|}
& $n$ & \multicolumn{3}{c|}{SIND} & \multicolumn{2}{c|}{SIHC}\\
\hline
 &  & $t_I$ & $t_A$ & $t_\sigma$ & $t$ & $t_\sigma$ \\
\hline
Do728 & $10^6$ & 30 & 0 & $1-3\sigma$ & 500 & $1-3\sigma$\\
A320  & $10^4$ & 20 & 0 & $1-3\sigma$ & 105 & $1-3\sigma$ \\
generic&$10^3$ & 20 & 0 & $1-3\sigma$ & 100 & $1-3\sigma$ \\
\end{tabular}
\end{ruledtabular}
\end{table}

Figure~\ref{fig:Figure04} shows the result of the procedure for the A320 with the selected thresholds given in Table~\ref{tab:Table01}. No sources are rejected as fitting artifacts ($t_A$=0). The crosses mark the determined centers of the sources, the numbers correspond to the order in which they are identified (descending $\hat{A}$). Figure~\ref{fig:Figure05} shows the result of the procedure for the Do728\deleted{ for $t_I=30$, no sources are rejected as fitting artifacts}. Finally, we calculate for all source-parts the probability $P$ of belonging to each source\deleted{-cluster} using $\text{PDF}_{S_i}$ and assign them to the source with the highest probability. Then we drop all source-parts with a PDF value below a threshold $t_\sigma$. Thus, each source-part is either assigned to a single source or classified as noise if $P(s\in S)<t_\sigma$. In Figure~\ref{fig:Figure04} and Figure~\ref{fig:Figure05}, the ellipses around the marked sources represent $\text{PDF}_S(\vec{x})=1-3\sigma$ and thus indicate the spatial locations (ROI) that are assigned to the corresponding sources.\\

\begin{figure}
	\centering
	\includegraphics[width=\reprintcolumnwidth]{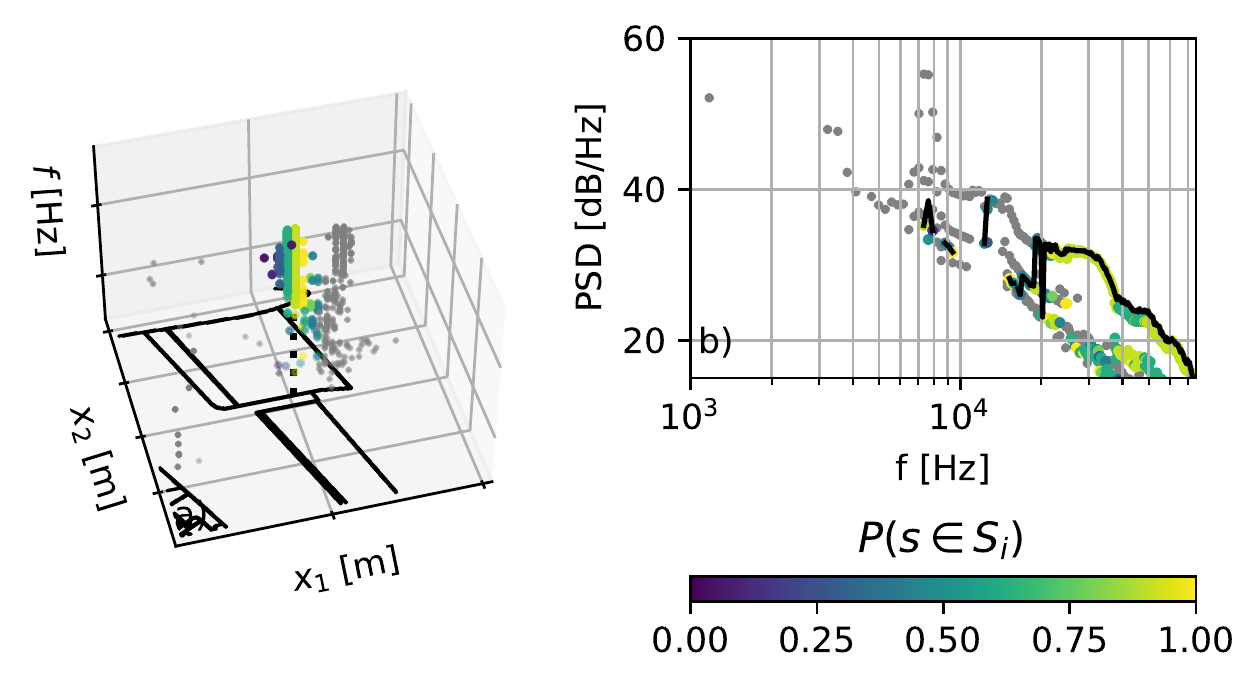}
	\caption{(Color online) $a)$ shows the source-parts of the A320 upstream flap side edge region (source number 2 in Figure~\ref{fig:Figure04}) at $M=0.175$, $\alpha=\SI{9}{\degree}$, $b)$ shows the same source-parts without the $x_i$-information. The color represents the source-parts' conditional probability of belonging to the source $P(s\in S_j)$ under the condition that they were assigned to it, gray source-parts were rejected as noise or assigned to another source. The black line represents the integrated spectrum from all source-parts that were assigned to the source.}
	\label{fig:Figure06}
\end{figure}

\begin{figure}
	\centering
	\includegraphics[width=\reprintcolumnwidth]{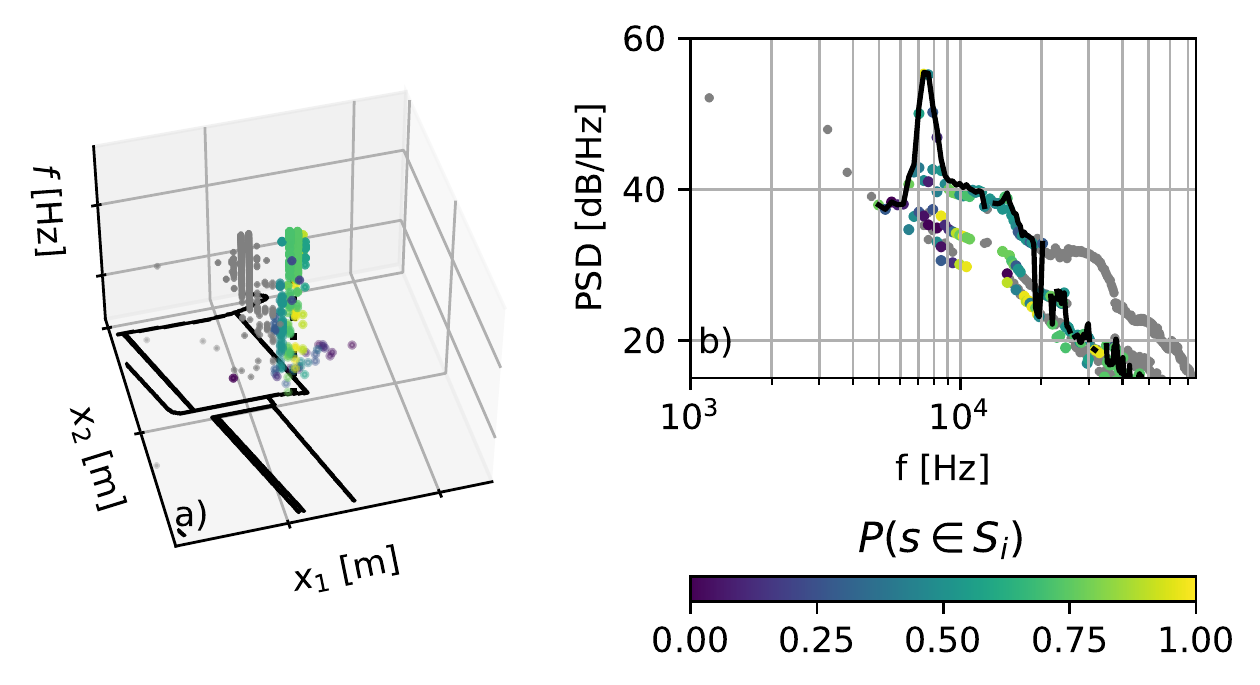}
	\caption{(Color online) The Figure shows the source-parts of the A320 downstream flap side edge region (source number 8 in Figure~\ref{fig:Figure04}) at $M=0.175$, $\alpha=\SI{9}{\degree}$, according to the description in Figure~\ref{fig:Figure06}.}
	\label{fig:Figure07}
\end{figure}

\begin{figure}
	\centering
	\includegraphics[width=\reprintcolumnwidth]{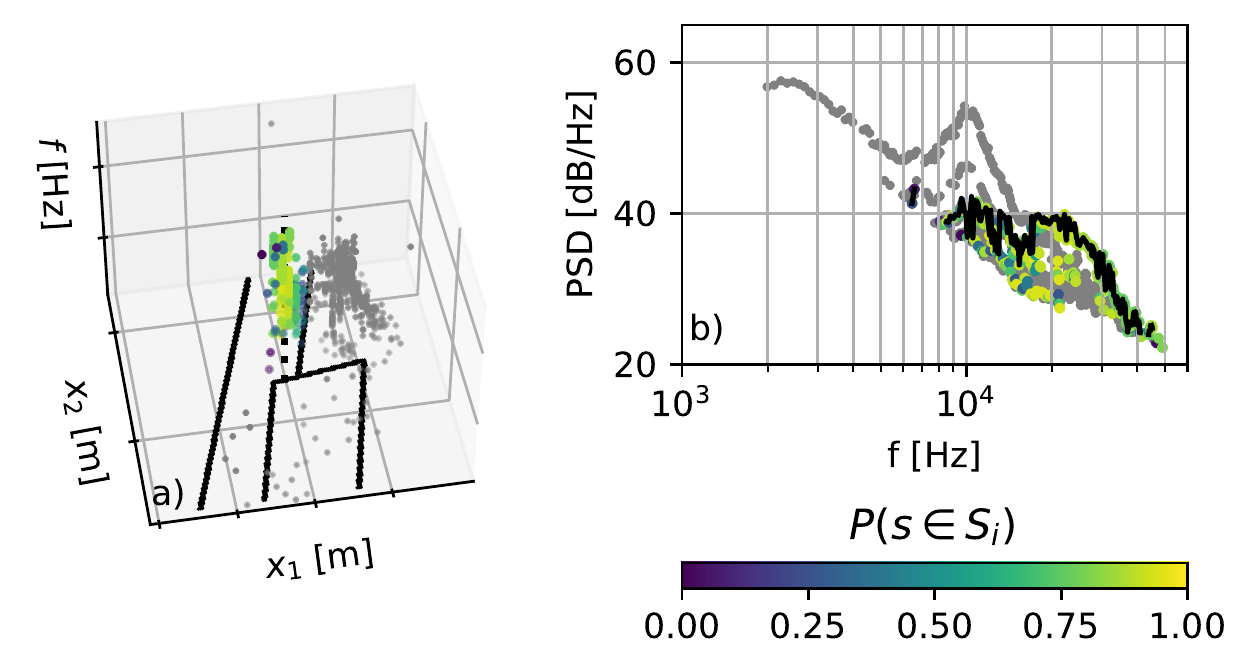}
	\caption{(Color online) $a)$ shows the source-parts of the Do728 upstream flap side edge region (source number 4 in Figure~\ref{fig:Figure05}) at $M=0.250$, $\alpha=\SI{6}{\degree}$, according to the description in Figure~\ref{fig:Figure06}.}
	\label{fig:Figure08}
\end{figure}

\begin{figure}
	\centering
	\includegraphics[width=\reprintcolumnwidth]{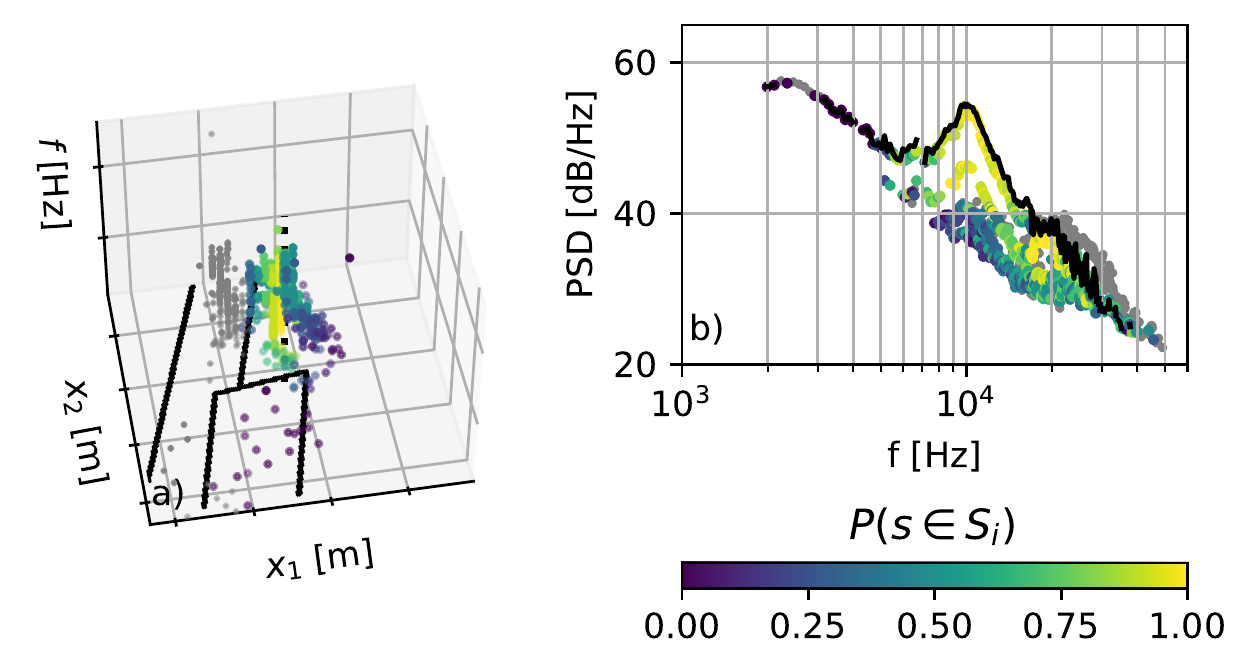}
	\caption{(Color online) $a)$ shows the source-parts of the Do728 downstream flap side edge region (source number 13 in Figure~\ref{fig:Figure05}) at $M=0.250$, $\alpha=\SI{6}{\degree}$, according to the description in Figure~\ref{fig:Figure06}.}
	\label{fig:Figure09}
\end{figure}

\begin{figure}
	\centering
	\includegraphics[width=\reprintcolumnwidth]{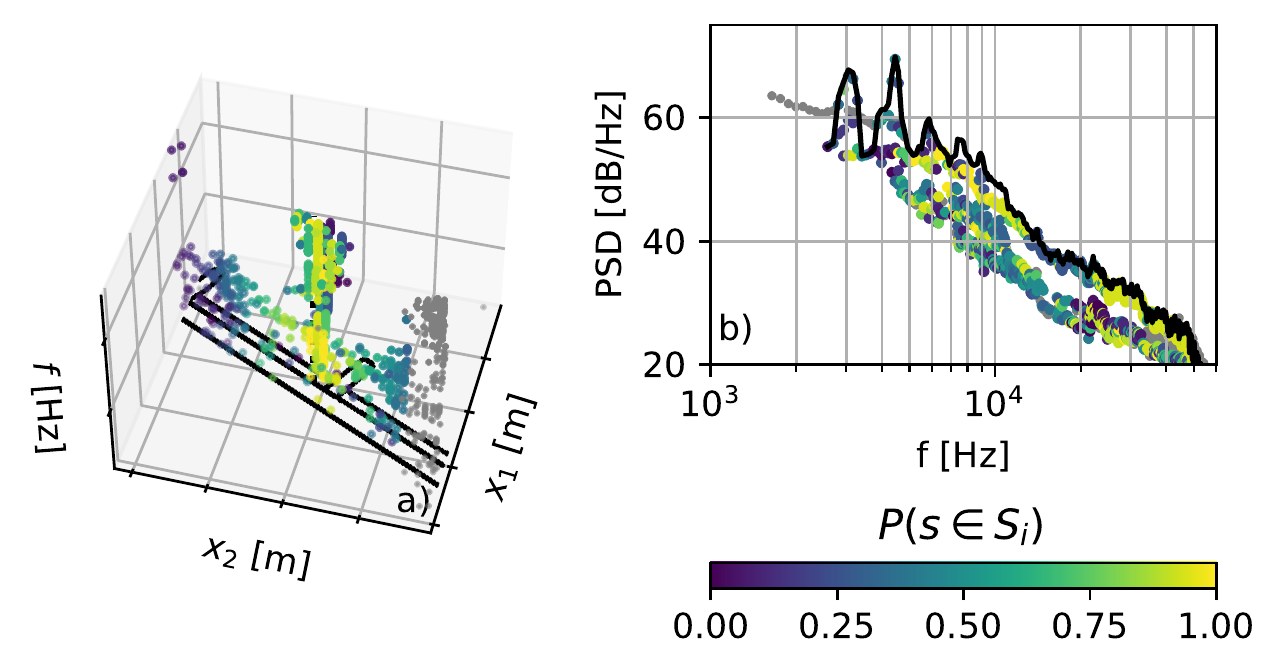}
	\caption{(Color online) $a)$ shows the source-parts at the DO728 slat track (source number 5 in Figure~\ref{fig:Figure05}) at $M=0.250$, $\alpha=\SI{1}{\degree}$, according to the description in Figure~\ref{fig:Figure06}.}
	\label{fig:Figure10}
\end{figure}

Figure~\ref{fig:Figure06} shows the methods result in detail for the leading flap side edge (LFSE) source location (see also Figure~\ref{fig:Figure03} for the DO728 LFSE fit, which shows two source-part distributions in this region) and all source-parts in this region. The source-parts' color encodes their corresponding PDF value. This can be interpreted as the conditional probability that they belong to the assigned source under the condition that they were assigned to it. Gray source-parts were either rejected as noise or assigned to another source, as its PDF (i.e. probability of belonging to this source) was higher in these spatial locations. Figure~\ref{fig:Figure06} $a)$ shows the source-part distribution on the 2D focus grid, the z-axis displays the frequency. Figure~\ref{fig:Figure06} $b)$ shows all source-parts from the region depicted in $a)$, neglecting the $x_i$-information. In Figure~\ref{fig:Figure06} $b)$ we observe multiple horizontal rows of points. They can either have a different shape, which indicates that these are the source-parts from two different sources or a similar shape with a simple vertical decibel offset. If the latter is observed, we assume that these rows at a lower PSD are artifacts from the CLEAN-SC process, as CLEAN-SC failed to remove these source-parts from the dirty map without residue. If the source-part rows have a different shape and are expected to be different source PSDs, the optimal scenario would be if one of them is assigned to the source with high confidence (bright color) and the other ones are rejected (gray color). Figure~\ref{fig:Figure07} shows the same for the downstream flap side edge region. The top row of source parts in the Figures shows, from low to high frequencies, a tonal peak around $f\approx\SI{6}{\kilo\hertz}$, and then three humps at $f\approx\SI{15}{\kilo\hertz}$, $f\approx\SI{30}{\kilo\hertz}$, and $f\approx\SI{50}{\kilo\hertz}$. Most of the source-parts of the first peak and hump were assigned to the TFSE, the source-parts of the two high-frequency humps were mostly assigned to the LFSE. A detailed analysis of how well this separation is performed is given in section~\ref{sec:comparision}. After integrating all source-parts that were assigned to the source over the frequency, we obtain the source spectra, indicated by the black line in Figure~\ref{fig:Figure06} and Figure~\ref{fig:Figure07}. In these examples, the spectra are mostly smooth but around $f\approx\SI{20}{\kilo\hertz}$ there are strong artifacts from incorrectly assigned source-parts. Figure~\ref{fig:Figure08} and Figure~\ref{fig:Figure09} show the corresponding results for the Do728 flap side edge. Figure~\ref{fig:Figure10} shows an exemplary Do728 slat / slat track source. This source will be analyzed in detail in section~\ref{sec:comparision}, since we assume it to be a complex spatial superposition of a line source (slat) and a point source (slat track).\\

\begin{figure}
	\centering
	\includegraphics[width=\reprintcolumnwidth]{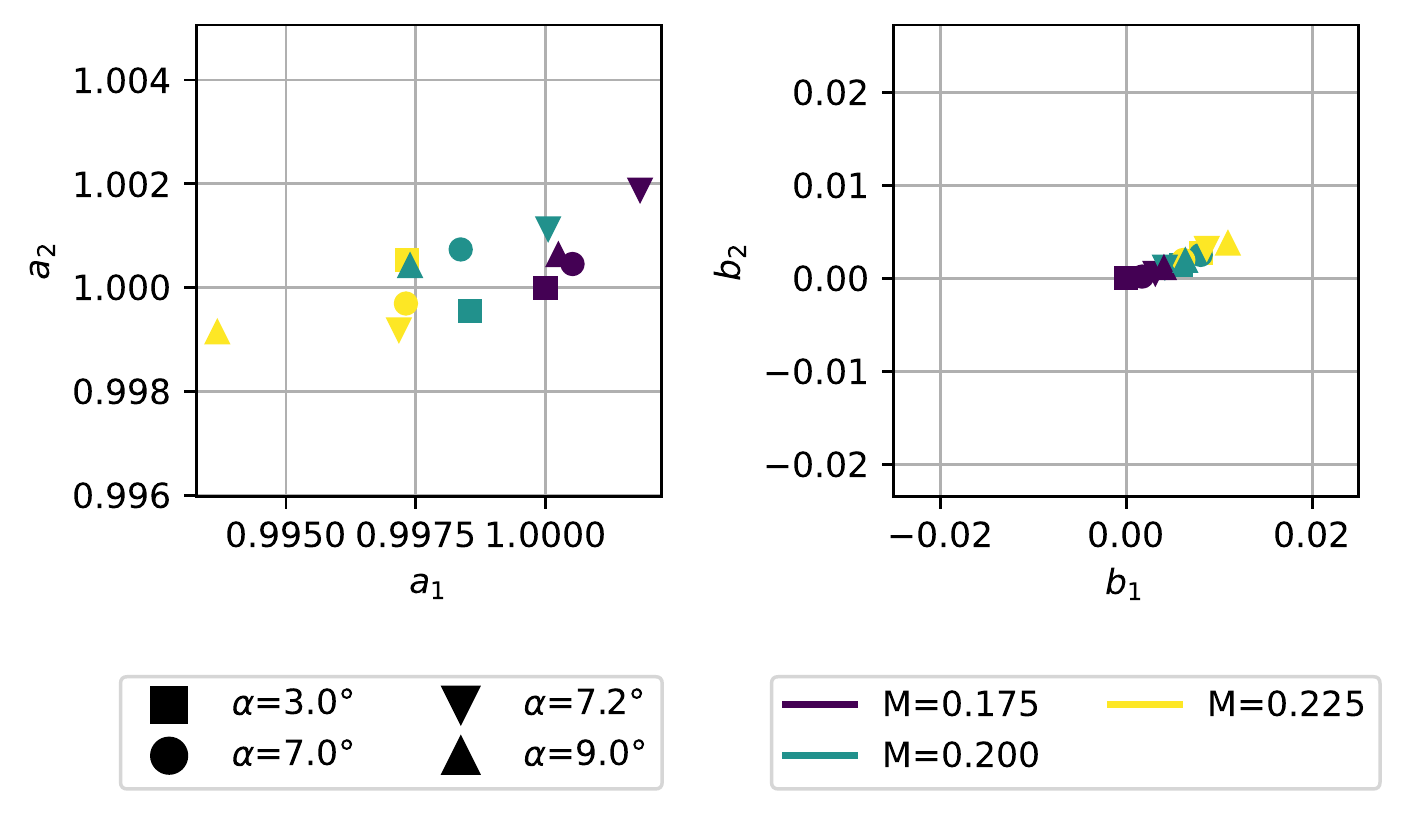}
	\caption{(Color online) A320, beamforming map alignment stretch $a_i$ and shift parameters $b_i$ for the source-part positions $x_i$ relative to the reference beamforming map at $M_1=0.175$, $\alpha_1=\SI{3}{\degree}$.}
	\label{fig:Figure11}
\end{figure}

SIND assumes that the source positions do not fundamentally change in the beamforming map over $M$ or $\alpha$ (considering a focus grid that rotates and moves with $\alpha$) so that the source-parts of different measurement configurations can be simply stacked and fitted at once to obtain global source positions and distributions, as shown in the results above. However, beamforming can suffer from the approximation of Greens Functions in complex medium flows to calculate the sound propagation from the source position to the microphone array or errors in the position of the focal plane\citep{Beamforming2019}. The first results in a shift or stretch of the beamforming maps, the second results in a source that moves through the map with increasing angle $\alpha$ because of the projection error (the strakes of the Do728 in Figure~\ref{fig:Figure05} show this behavior). The first problem can be fixed by aligning the beamforming maps prior to fitting the normal distributions. To do so, the source-part histogram of each individual configuration is calculated, then a histogram is chosen as a reference. All remaining histogram positions are then linearly modified with
\begin{equation}\label{eq:map_alignment}
    f(x_i)=a_ix_i+b_i
\end{equation}
to achieve a maximum 2D spatial correlation with the reference histogram using standard optimization methods. Eq.~\ref{eq:map_alignment} is then used to modify the source-parts' positions $x_i$ prior to calculating the global histogram. Figure~\ref{fig:Figure11} shows the obtained parameters $a_i, b_i$ for the A320. While the stretch factors $a_i$ are small, the shift factors $b_i$ show a clear trend. The beamforming maps shift slightly with increasing angle of attack and substantially with increasing Mach number downstream (more than $b_1\ge2\Delta x_1$).

\subsection{Source Identification based on Hierarchical Clustering (SIHC)}
\begin{figure}
	\centering
	\includegraphics[width=\reprintcolumnwidth]{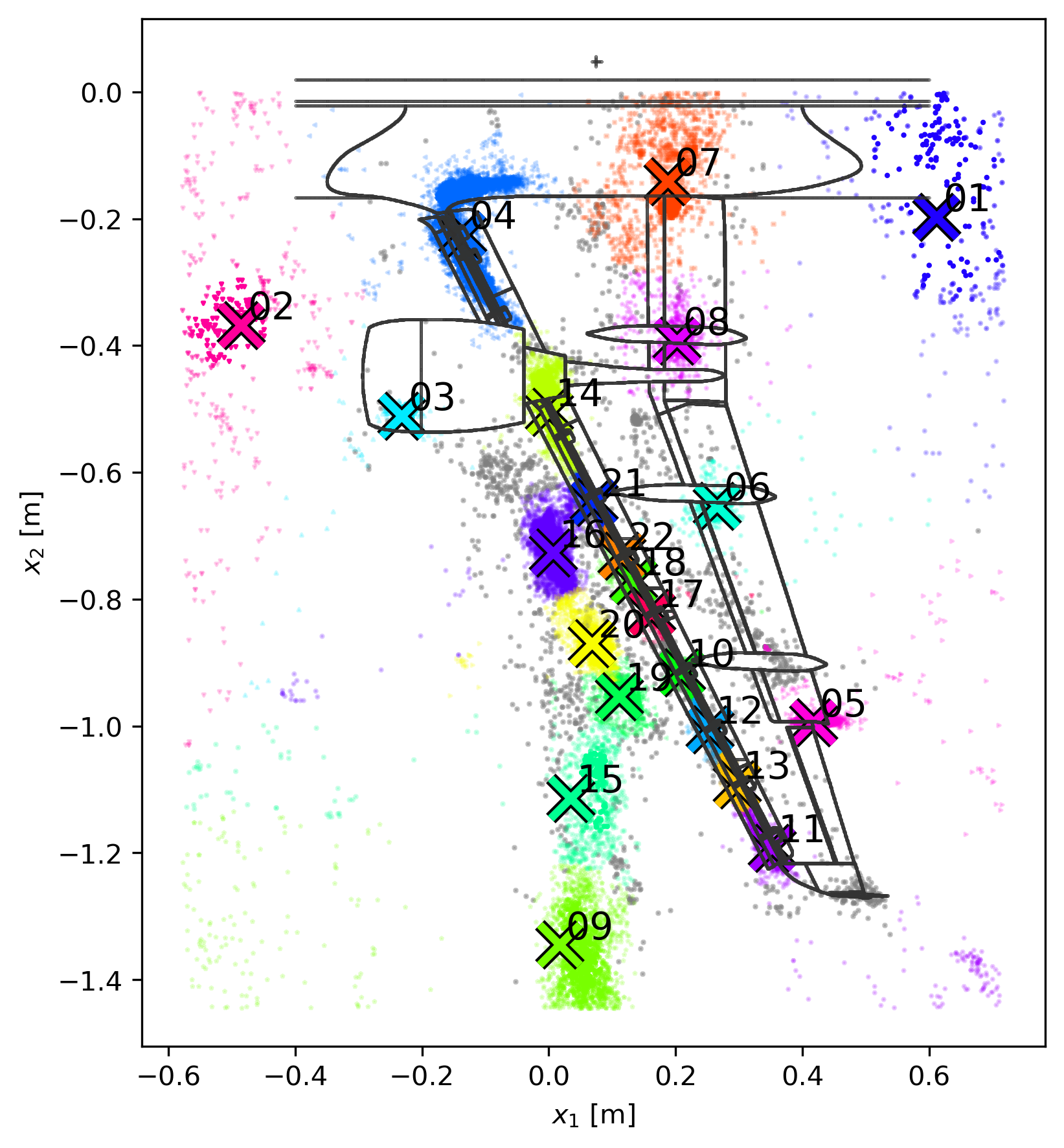}
	\caption{(Color online) A320. Resulting clusters from HDBSCAN at $t=105$, using an euclidean distance metric. The cluster midpoints are marked, the corresponding source-parts are displayed in the same color. The color intensity displays the probability of belonging to the cluster. Gray source-parts were rejected as noise.}
	\label{fig:Figure12}
\end{figure}

\begin{figure}
	\centering
	\includegraphics[width=\reprintcolumnwidth]{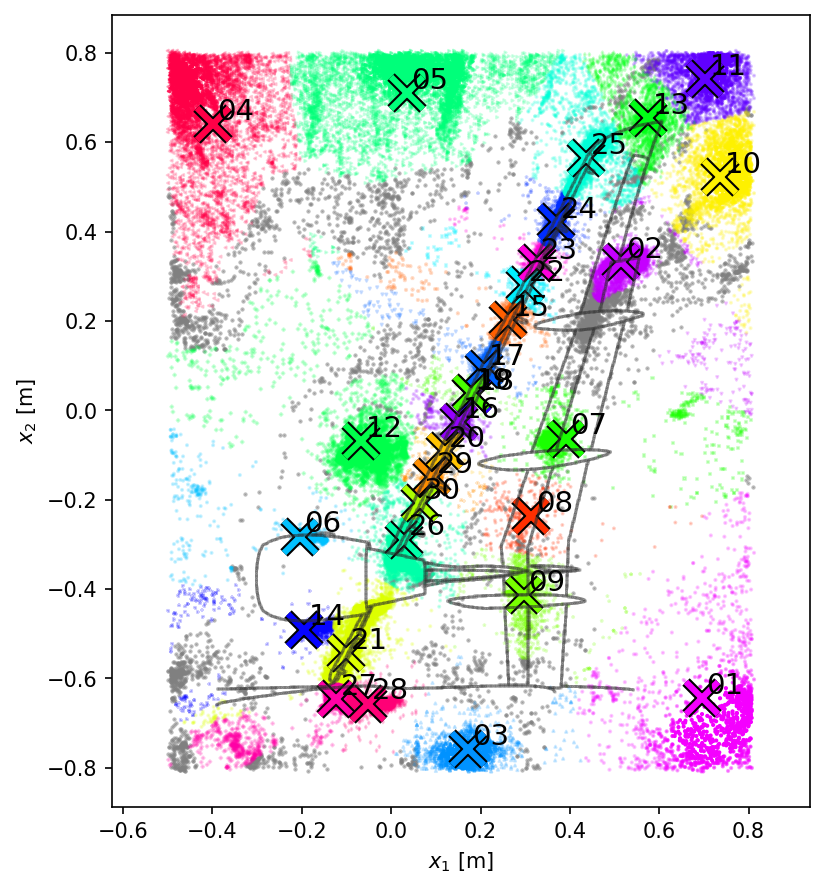}
	\caption{(Color online) Do728. Resulting clusters from HDBSCAN at $t=500$, using an euclidean distance metric. The cluster midpoints are marked, the corresponding source-parts are displayed in the same color. The color intensity displays the probability of belonging to the cluster. Gray source-parts were rejected as noise.}
	\label{fig:Figure13}
\end{figure}

A second approach to identifying sources and assigning the corresponding source-parts is clustering methods which can automatically group source-parts in a multidimensional space. Since we do not know the number of expected clusters (sources) and their distribution beforehand, we choose Hierarchical Density-Based Spatial Clustering of Applications with Noise~\citep{Campello2013,McInnes2017} (HDBSCAN). Similar to SIND, HDBSCAN requires a threshold $t$ below which a cluster is rejected as noise. The threshold has a great effect on the resulting clusters and has to be determined with the expert in the loop. We cluster the source-parts based on their normalized location $\vec{x}_i$, normalized Strouhal number $\text{St}_i$ and Mach scaled, normalized PSD level (normalized to the range $[0,1]$), thus in 4D-space. When clustering source-parts of maps at different Mach numbers at the same time, we recommend using a Mach scaled PSD 
\begin{equation}\label{eq:Mach_scaling}
\widehat{\text{PSD}}=\text{PSD}-10\log{M^n}
\end{equation}
with $n\approx5.5$ and a normalized frequency like the Strouhal or Helmholtz number. This scaling ensures that the source-parts of sources at different Mach numbers are roughly at the same location in the frequency and PSD-level space, as aeroacoustic noise generally scales around this Mach exponent~\citep{Guo2003}.\\

\begin{figure}
	\centering
	\includegraphics[width=\reprintcolumnwidth]{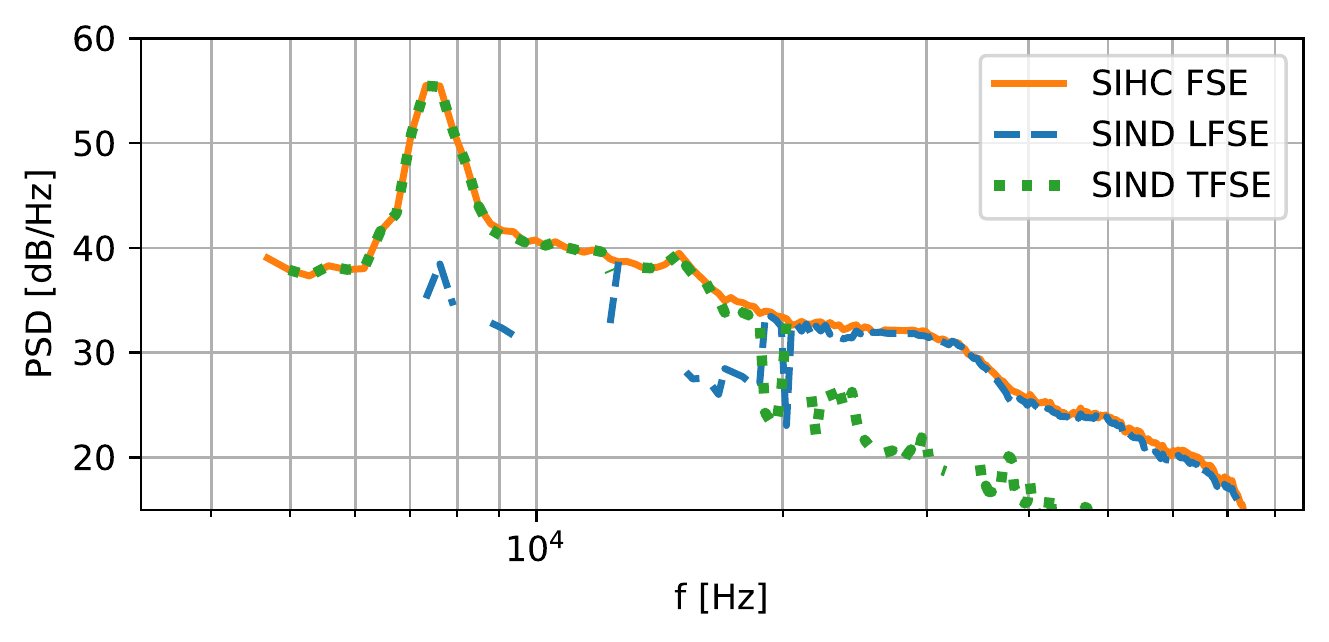}
	\caption{(Color online) Comparison of the resulting source spectra from the SIND and SIHC method at $M=0.175$, $\alpha=\SI{9}{\degree}$ for the A320 flap side edge (FSE), respectively leading flap side edge (LFSE) and trailing flap side edge (TFSE).}
	\label{fig:Figure14}
\end{figure}

\begin{figure}
	\centering
	\includegraphics[width=\reprintcolumnwidth]{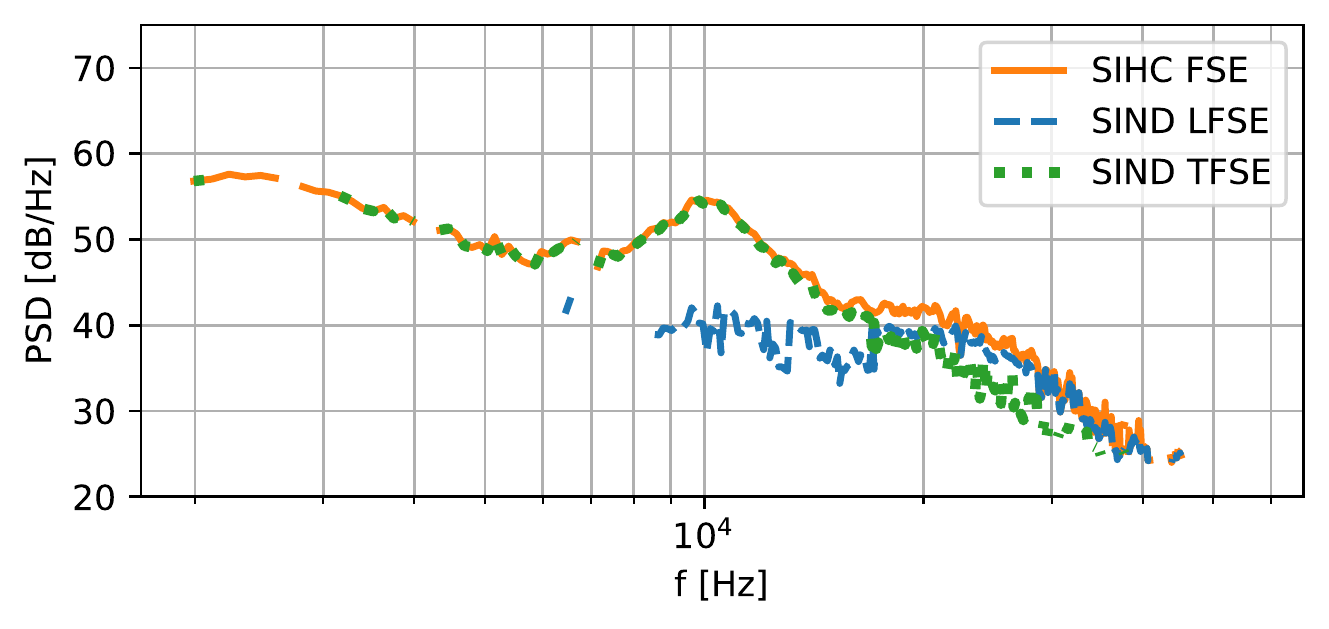}
	\caption{(Color online) Comparison of the resulting source spectra from the SIND and SIHC method at $M=0.125$, $\alpha=\SI{7}{\degree}$ for the Do728 flap side edge (FSE), respectively leading flap side edge (LFSE) and trailing flap side edge (TFSE).}
	\label{fig:Figure15}
\end{figure}

\begin{figure}
	\centering
	\includegraphics[width=\reprintcolumnwidth]{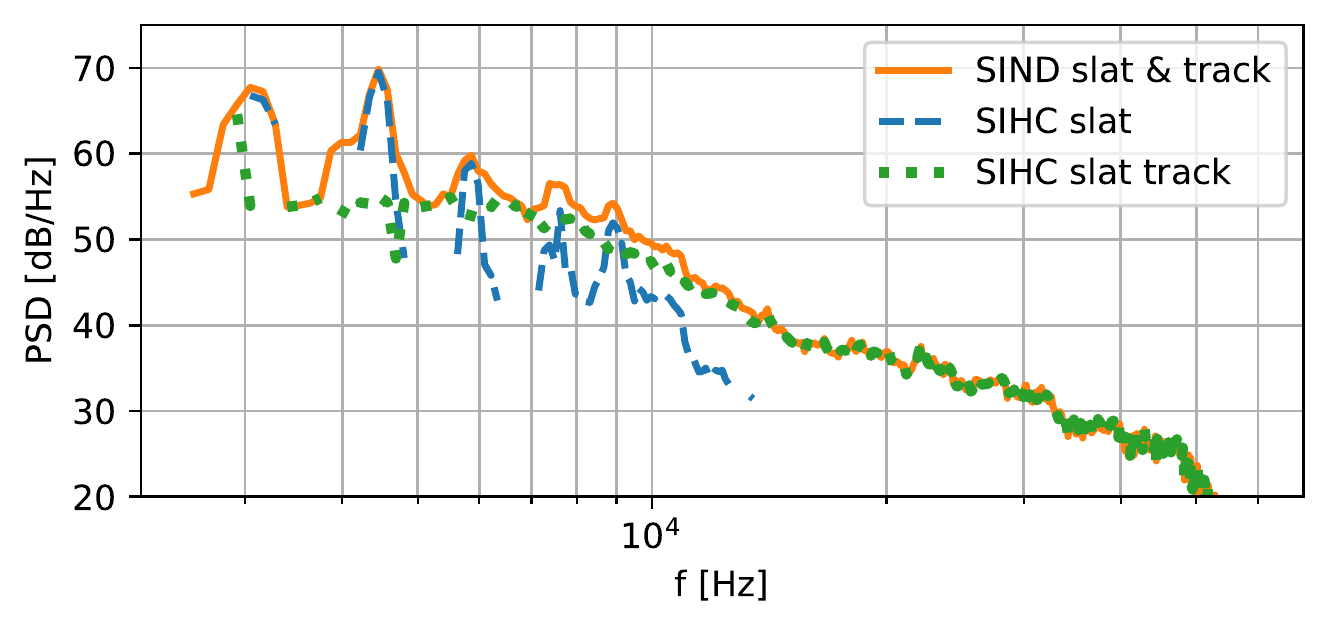}
	\caption{(Color online) Comparison of the resulting source spectra from the SIND and SIHC method at $M=0.250$, $\alpha=\SI{1}{\degree}$ for the Do728 slat, the slat track and the combined SIND ROI.}
	\label{fig:Figure16}
\end{figure}

Figure~\ref{fig:Figure12} shows the result of HDBSCAN for the A320 at $t=105$ and Figure~\ref{fig:Figure13} for the Do728 at $t=500$\added[R2C07]{, see Table~\ref{tab:Table01}}. The crosses mark the cluster midpoints of the corresponding source-parts, displayed in the same color. Gray source-parts are rejected as noise as their confidence of belonging to any source is below $t_\sigma=1-3\sigma$. The color intensity displays the classification confidence. \replaced{Figure~\ref{fig:Figure14} shows the A320 source-parts that were assigned to the flap side edge region at $M=0.125$, $\alpha=\SI{9}{\degree}$ within $t_\sigma=1-3\sigma$ confidence}{Figure~\ref{fig:Figure14} shows the resulting integrated spectra from the A320 flap side edge region in comparison to the SIND method}, Figure~\ref{fig:Figure15} shows the same for the Do728. Figure~\ref{fig:Figure16} shows the same slat track source from the SIND solution in Figure~\ref{fig:Figure10} \replaced{and Figure~\ref{fig:Figure16} shows}{as well as} the upper part of the corresponding slat \added{for the SIHC solution}.\\

\subsection{Comparison of SIND and SIHC}\label{sec:comparision}

\begin{figure}
	\centering
    \includegraphics[width=\reprintcolumnwidth]{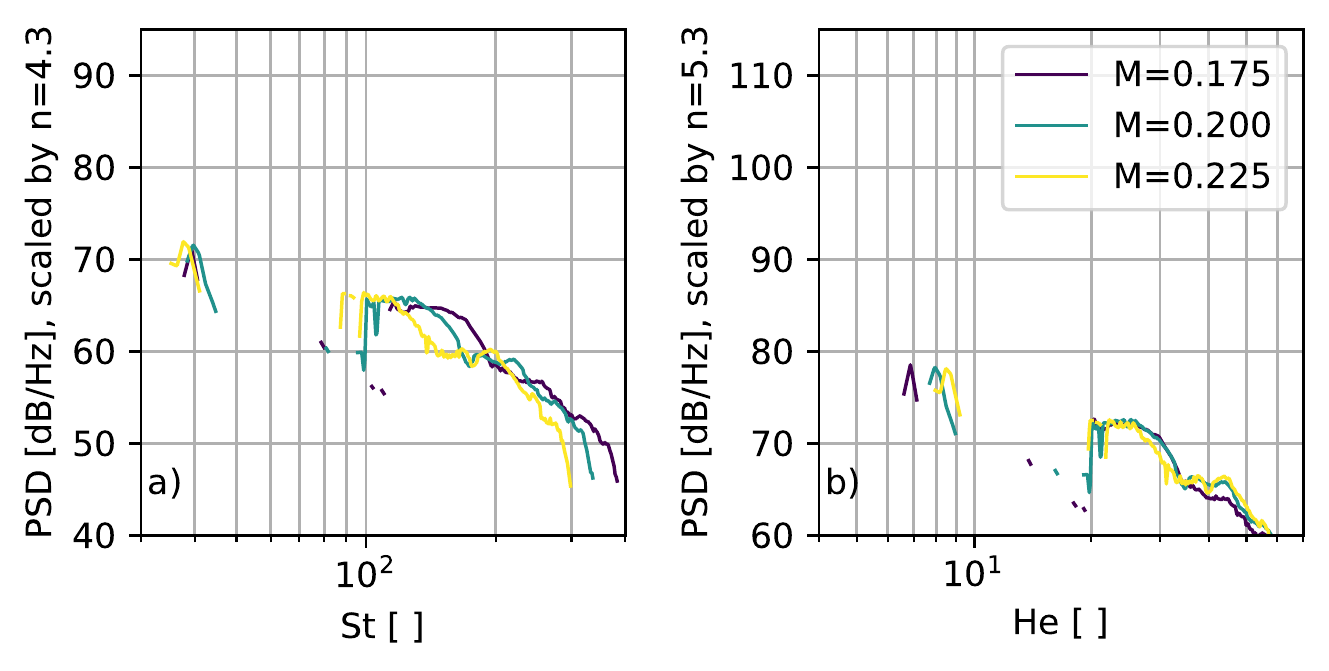}
	\caption{(Color online) The A320 spectra of the SIND leading flap side edge source (number 2 in Figure~\ref{fig:Figure04}) at $\alpha=\SI{9}{\degree}$ over $a)$ Strouhal number and $b)$ Helmholtz number. The spectra are Mach scaled with the scaling exponent $n$, see eq.~\ref{eq:Mach_scaling}.}
	\label{fig:Figure17}
\end{figure}

\begin{figure}
	\centering
    \includegraphics[width=\reprintcolumnwidth]{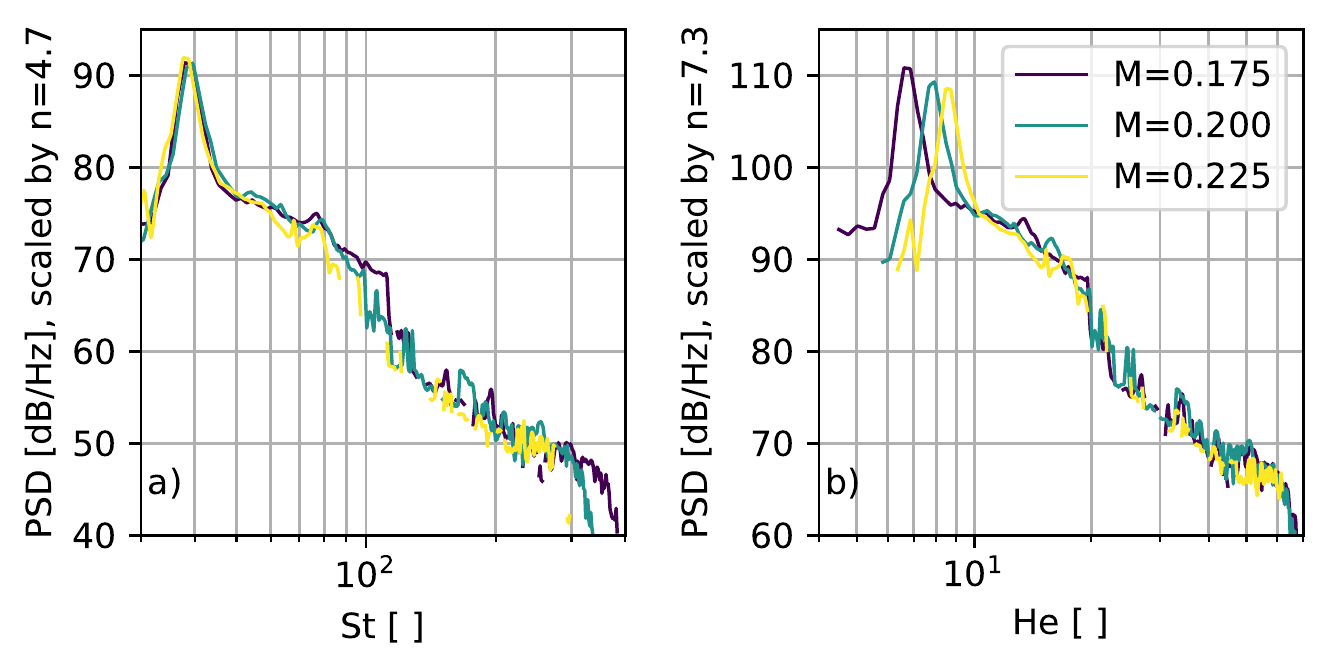}
	\caption{(Color online) The A320 spectra of the SIND trailing flap side edge source (number 8 in Figure~\ref{fig:Figure04}) at $\alpha=\SI{9}{\degree}$ over $a)$ Strouhal number and $b)$ Helmholtz number. The spectra are Mach scaled with the scaling exponent $n$, see eq.~\ref{eq:Mach_scaling}.}
	\label{fig:Figure18}
\end{figure}

\begin{figure}
	\centering
    \includegraphics[width=\reprintcolumnwidth]{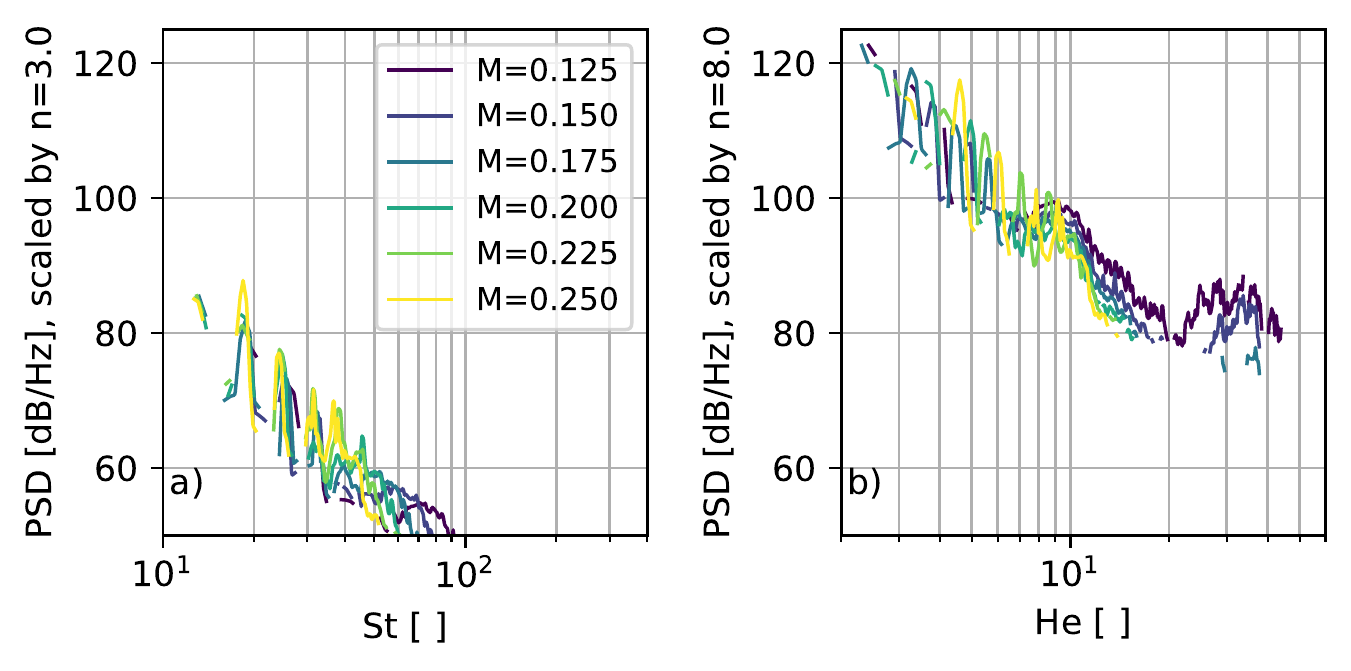}
	\caption{(Color online) The Do728 spectra of the SIHC slat source (number 20 in Figure~\ref{fig:Figure13}) at $\alpha=\SI{1}{\degree}$ over $a)$ Strouhal number and $b)$ Helmholtz number. The spectra are Mach scaled with the scaling exponent $n$, see eq.~\ref{eq:Mach_scaling}.}
	\label{fig:Figure19}
\end{figure}

\begin{figure}
	\centering
    \includegraphics[width=\reprintcolumnwidth]{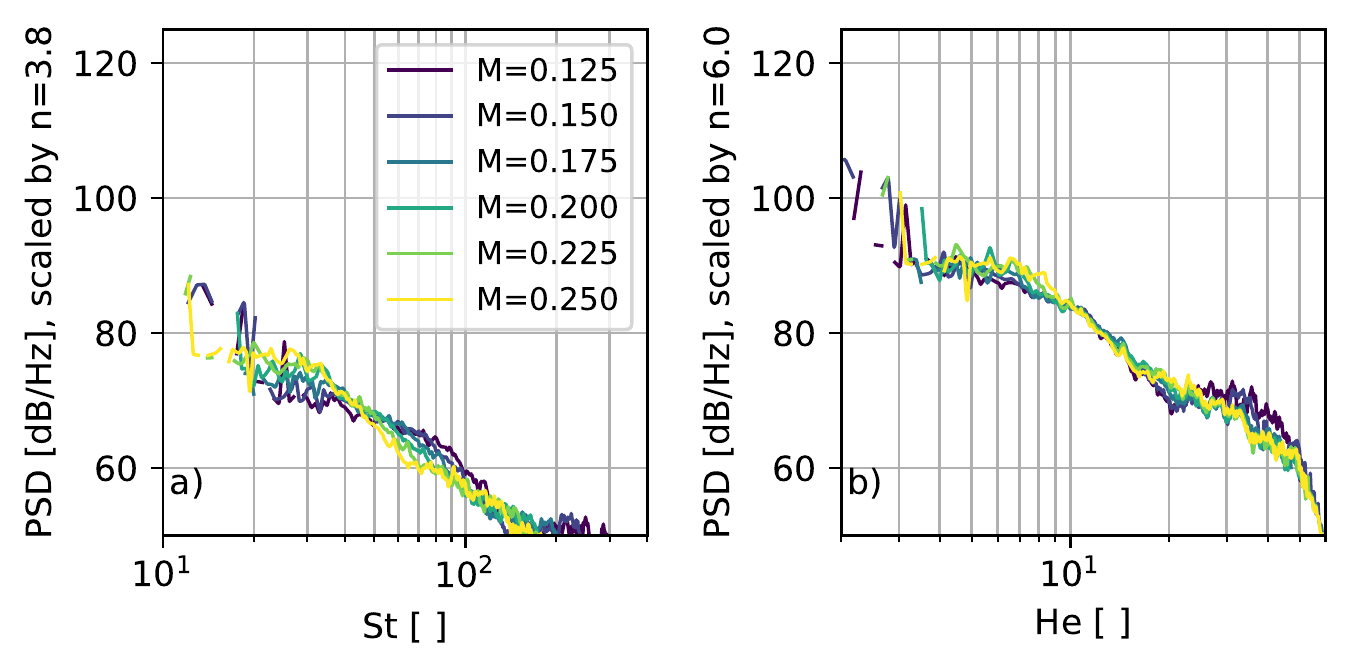}
	\caption{(Color online) The Do728 spectra of the SIHC slat track source (number 29 in Figure~\ref{fig:Figure13}) at $\alpha=\SI{1}{\degree}$ over $a)$ Strouhal number and $b)$ Helmholtz number. The spectra are Mach scaled with the scaling exponent $n$, see eq.~\ref{eq:Mach_scaling}.}
	\label{fig:Figure20}
\end{figure}
To assess the quality of the ROIs, both methods are compared to each other and the authors' expectations. Both methods yield comparable ROIs and are able to identify the prominent source locations such as the flap side edge, slat tracks, wingtip, or strakes that are indicated by the blobs in the corresponding histograms in Figure~\ref{fig:Figure04} and Figure~\ref{fig:Figure05}. SIND often separates individual sources in dense and overlapping source regions that are clustered together by SIHC, especially at the inner slat or the flap side edge region. SIHC finds additional source regions that are not well localized and spread over the map, especially sources that are not located on the wing, such as what we assume to be wind-tunnel noise reflections. We observe that SIND and SIHC often find comparable sub- or super-sources in the sense that some sources detected in SIND correspond to multiple sources detected by SIHC or vice-versa, e.g. the flap side edge in Figure~\ref{fig:Figure14} and Figure~\ref{fig:Figure15} or the slat / track in Figure~\ref{fig:Figure16} for the Do728. To assess the quality and legitimacy of the ROI separation, a self-similarity analysis is performed. Thus, the spectra levels are power scaled with eq.~\ref{eq:Mach_scaling} and displayed over Strouhal and Helmholtz number. While a self-similarity across the whole spectrum does not necessarily imply that the whole spectrum is generated by the same mechanism, a self-similarity over multiple frequency types in different frequency intervals doubtlessly shows this~\citep{Mueller1979}.\\

For the A320 flap side edge, a self-similarity analysis shows that the up- and downstream separation of SIND is reasonable, see Figure~\ref{fig:Figure17} and Figure~\ref{fig:Figure18}. While the low-frequency peak scales over Strouhal number, the high-frequency humps scale over Helmholtz number which suggests different aeroacoustic source mechanisms and justifies the spatial separation. Dobrzynski points out that the complex acoustical behavior of the flap side edge is a combination of trailing-edge noise, noise of a primary suction side vortex, a secondary suction side vortex, their mixing and accelerated free turbulence in the vortex flow~\citep{Dobrzynski2010}, which supports this result. We explicitly see in Figure~\ref{fig:Figure18} that the smaller, high-frequency hump is also self-similar over the Strouhal number which indicates that it is assigned to the correct source. The analysis of the Do728 flap side edge shows the same self-similarities (not shown). While SIND and SIHC separate most slat and slat tracks, SIHC reconstructs more smooth spectra than SIND by correctly identifying the corresponding source-parts. Figure~\ref{fig:Figure10} shows, that the low-frequency slat tones are not well localized and scattered around the slat area, which matches Dobrzynski's hypothesis, that these tones result from model-scale low Reynolds numbers and are generated by coherent laminar flow separation at the slat hook and thus, are line sources~\citep{Dobrzynski2001, Dobrzynski2010}. By distribution assumption, SIND assumes point-like sources, which cannot properly detect these line-sources. Even if so, SIND only assigns the source-parts based on their spatial distribution to the sources, but these sources spatially overlap. SIHC on the other hand not only separates the Strouhal number scaling slat tones, see Figure~\ref{fig:Figure19}, from the Helmholtz number scaling slat track source, see Figure~\ref{fig:Figure20}, it assigns the source-parts mostly correct in terms of self-similar behavior to the corresponding source spectra. This is possible due to the additional frequency and SPL information, based on which the clusters are identified.\\

Performance-wise SIHC's computation time scales around $\mathbf{O}(n\log n)$ for the number $n$ of source-parts~\citep{McInnes2017}. Since SIND does not cluster the points directly, the computation time is independent of the number of points, which is a huge advantage for large datasets. The total number of source-parts in the Do728 dataset is around $n=10^6$, which SIND processes within seconds and SIHC within an hour on a standard laptop. Both methods process the A320 dataset within seconds, which contains around $n=10^4$ source-parts.

\section{Method errors}
As stated in the sections above, it is not possible to quantitatively estimate the methods' errors on the real-world datasets due to the lack of a ground truth. Thus, a generic test with a streamlined monopole-loudspeaker in an open wind-tunnel is used to validate the methods, give an estimation of how well the source spectra are reconstructed, and how well the source positions are estimated. The loudspeaker was measured at three different positions $\vec{x}_1=[-0.055,0.105,0],\vec{x}_2=[0.105,0.105,0],\vec{x}_3=[0.255,0.105,0]$, with three different band-limited white noise signals, see Table~\ref{tab:Table02}. Additionally, a measurement with no speaker signal was performed at each configuration to obtain a noise-floor CSM that can be subtracted to reduce the noise of the wind-tunnel and speaker housing in the flow~\citep{Bahr2015}. A ground truth and error metric for the source-separation problem is achieved as follows.\\

\begin{table}
\caption{\label{tab:Table02}Absolute positional errors $|\varphi|$ in degree and absolute spectrum reconstruction errors $|\varepsilon|$ in decibel of SIND and SIHC performance and the integration of the individual CLEAN-SC source-maps (CLEAN). The first column lists the upper and lower butterworth band-limit frequencies $f$ in Hertz, the second column lists the corresponding filter roll-offs $r$ in decibel per octave for the three white noise sources.}
\begin{ruledtabular}
\begin{tabular}{c|cc||c|c|c|c|}
& $f$ & $r$ & & SIND & SIHC & CLEAN\\
\hline
\multirow{2}{*}{\begin{sideways}$S_1$\end{sideways}} &32k&48& $|\varphi|$  & $0.48\pm0.01$ &  $0.46\pm0.02$& $0.61\pm0.32$\\
                                                     &15k&24& $|\varepsilon|$ & $1.62\pm 2.35$ &  $1.64\pm 2.71$& $1.59\pm1.72 $\\
\hline
\multirow{2}{*}{\begin{sideways}$S_2$\end{sideways}} &32k&48& $|\varphi|$  & $0.28\pm0.06$ &  $0.28\pm0.06$&$0.45\pm0.92$\\
                                                     &20&48& $|\varepsilon|$ &  $2.19\pm 3.00$ & $2.19\pm 3.00$& $1.13\pm1.16 $\\
\hline
\multirow{2}{*}{\begin{sideways}$S_3$\end{sideways}} &5k&24& $|\varphi|$  & $0.28\pm0.11$ &  $0.43\pm0.20$&$0.28\pm0.10$\\
                                                     &20&48& $|\varepsilon|$ & $2.73\pm 7.34$ &  $3.45\pm7.88 $& $2.47\pm2.97 $\\
\hline
\multirow{3}{*}{\begin{sideways}total\end{sideways}} &&& $|\varphi|$  & $0.35\pm0.09$ &  $0.39\pm0.08$ & $0.45\pm1.13$\\
                                                     &&    & $|\varepsilon|$ &  $2.03\pm 3.66$ &  $2.14\pm3.99 $ &  $1.73\pm2.16 $\\
                                                     && & $f_r$ & \SI{65.6}{\percent} & \SI{66.8}{\percent}&\SI{96.8}{\percent}\\
\end{tabular}
\end{ruledtabular}
\end{table}

\begin{figure}
	\centering
    \includegraphics[width=\reprintcolumnwidth]{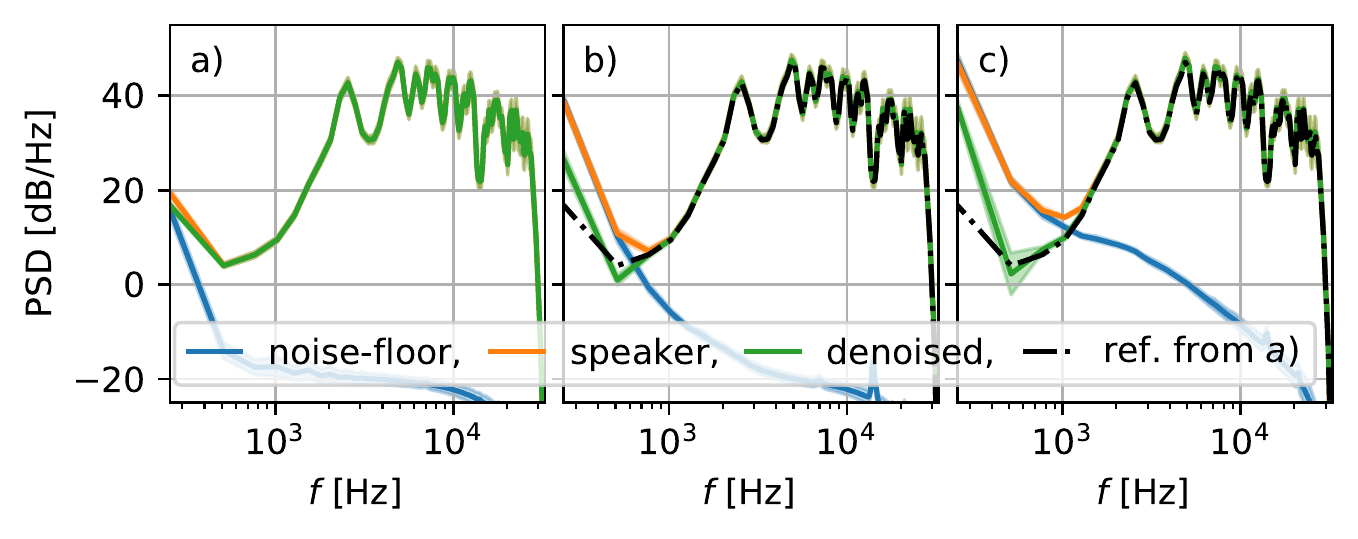}
	\caption{(Color online) CSM$_{i,i}$ auto-spectra of source $S_2$ at $a)$ $M=0.00$, $b)$ at $M=0.03$, and $c)$ at $M=0.06$. The blue lines represent the noise-floor with the speaker turned off, the orange lines represent the measurement with the speaker turned on, and the green lines represent the denoised measurement. The black lines in $b)$ and $c)$ show the denoised measurement (green line) from $a)$ as a comparison. The shaded areas depict the standard deviation over all microphones.}
	\label{fig:Figure21}
\end{figure}

First, the CSM auto-power spectra are compared for a single speaker position (source $S_2$) in Figure~\ref{fig:Figure21} with (orange line) and without (blue line) a noise signal. The denoised signal (green line) is achieved by subtracting the noise-floor CSM from the speaker signal CSM~\citep{Bahr2015}. It is observed that the flow-effects are neglectable on the radiated sound of the speaker at high frequencies. However, at low frequencies with a negative SNR (the wind-tunnel noise is louder than the speaker), the denoised signal still overestimates the PSD below $f\le\SI{500}{\hertz}$. Thus, the cleaned measurements of the individual sources at $M_1=0.00$ are regarded as the true immission levels for all Mach numbers. Then the speaker's reference emission level (i.e. sound power) is estimated by multiplying the true immission levels (at $M_1=0.00$) with the inverse Green's Function of a monopole, which equals to $\Delta L = 10\log_{10}(r_m)$, where $r_m=|\vec{x}_m-\vec{x}_s|$ is the distance between the fix source position $\vec{x}_s$ and each microphone position $\vec{x}_m$. This projected, microphone $m$ averaged emission level will be regarded as the ground truth sound power $\langle\text{PSD}_\text{true}\rangle_m$ with the Kronecker delta $\delta$
\begin{equation}\label{eq:ground_truth}
    \text{PSD}_{S_i,\text{true}} = \langle \delta_{m,n}\text{CSM}_{S_i,n}+20\log_{10}(|\vec{x}_m-\vec{x}_{S_i}|)\rangle_m \,.
\end{equation}

\begin{figure}
	\centering
    \includegraphics[width=\reprintcolumnwidth]{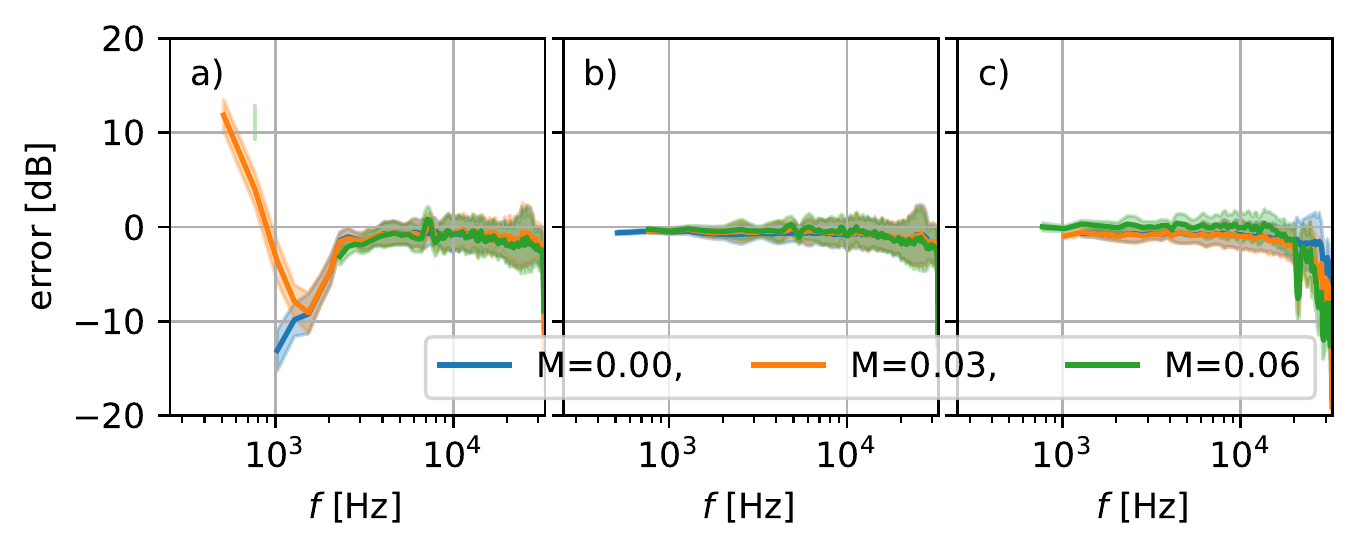}
	\caption{(Color online) Figure shows the error $\varepsilon=\text{PSD}_\text{CLEAN-SC} - \text{PSD}_\text{true}$, see eq.~\ref{eq:ground_truth}, for each Mach number for $a)$ $S_1$, $b)$ $S_2$, and $c)$ $S_3$. The shaded area depicts the standard deviation of the ground truth sound power over all microphones.}
	\label{fig:Figure22}
\end{figure}

Second, conventional beamforming and CLEAN-SC are performed on the individual, denoised source CSMs. This allows us to obtain an estimation of how well the individual source powers obtained by CLEAN-SC correspond to the ground truth. To obtain CLEAN-SC reference spectra from the beamforming maps, all source-parts within a spatial radius $r=\SI{0.1}{\metre}$ of the true source positions were integrated. Additionally, these source-parts' positions were averaged to obtain a source position estimation of the CLEAN-SC process. Figure~\ref{fig:Figure22} shows the error $\varepsilon=\text{PSD}_\text{CLEAN-SC} - \text{PSD}_\text{true}$ for all each individual source and all Mach numbers. The standard deviation depicts the variance over the microphone averaged ground-truth, see eq.~\ref{eq:ground_truth}. A cut-on frequency can be observed, below which CLEAN-SC is not able to reconstruct the sound source correctly. For source 1, below $f<\SI{1.5}{\kilo\hertz}$, the beamforming results massively over- or under-predict the PSD. The reason for this might be the insufficient cleaning of the CSM, as shown in Figure~\ref{fig:Figure21}, and the low SNR at these frequency intervals. The averaged position errors $|\varphi|$ and absolute spectra errors $|\varepsilon|$ of the CLEAN-SC process are given in Table~\ref{tab:Table02}. The position errors are given as angular errors with respect to the microphone array center instead of $\Delta x_i$, since beamforming localization usually depends on the distance of the focal plane. For comparison, the focus point resolution is $\Delta x_i=\SI{0.005}{\metre}\approx\SI{0.44}{\degree}$ in the center of the focus grid.\\

\begin{figure}
	\centering
    \includegraphics[width=0.9\reprintcolumnwidth]{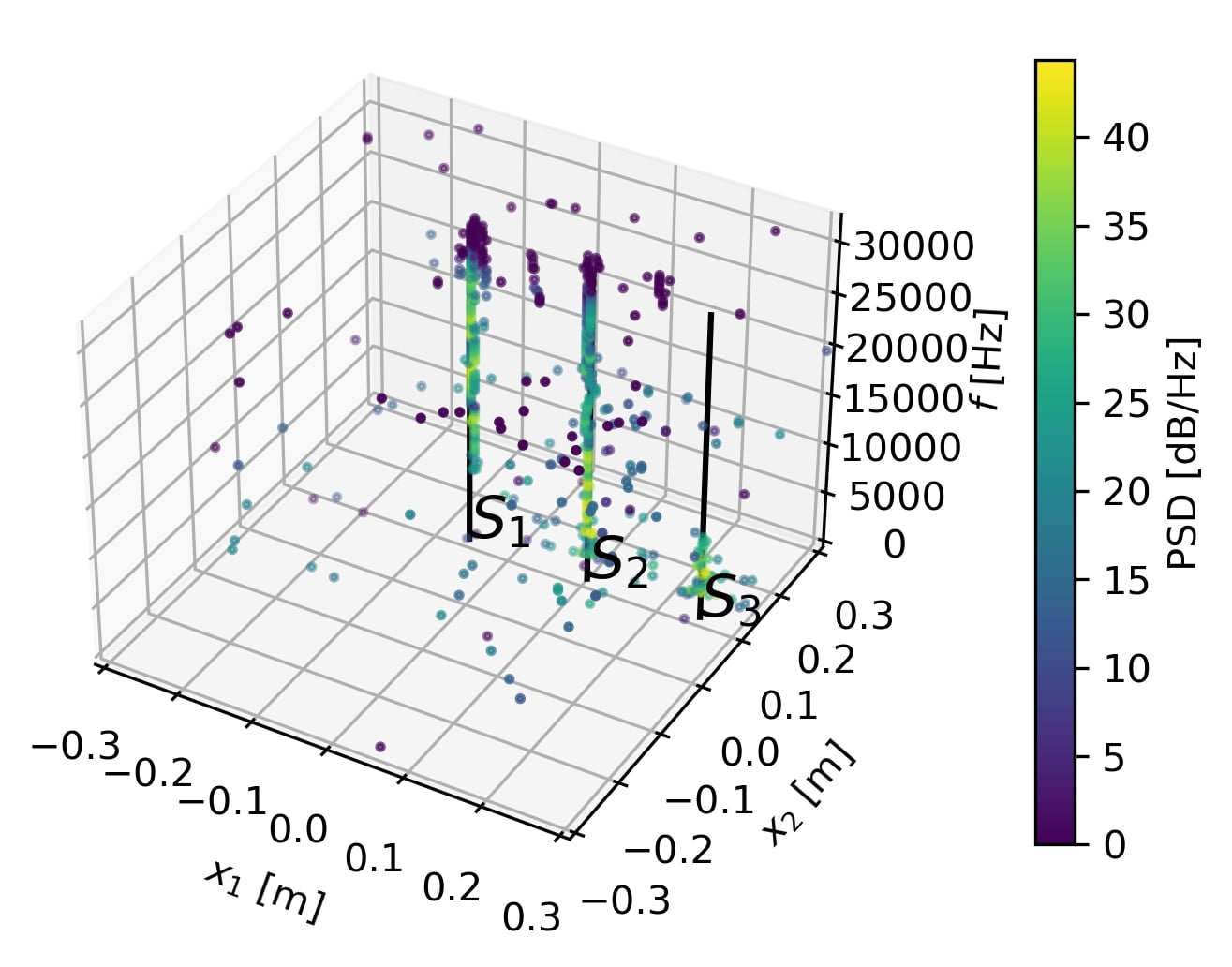}
	\caption{(Color online) CLEAN-SC result of the superpositioned, denoised CSMs at $M_3=0.06$. The true positions are marked with black lines.}
	\label{fig:Figure23}
\end{figure}

\begin{figure}
	\centering
    \includegraphics[width=\reprintcolumnwidth]{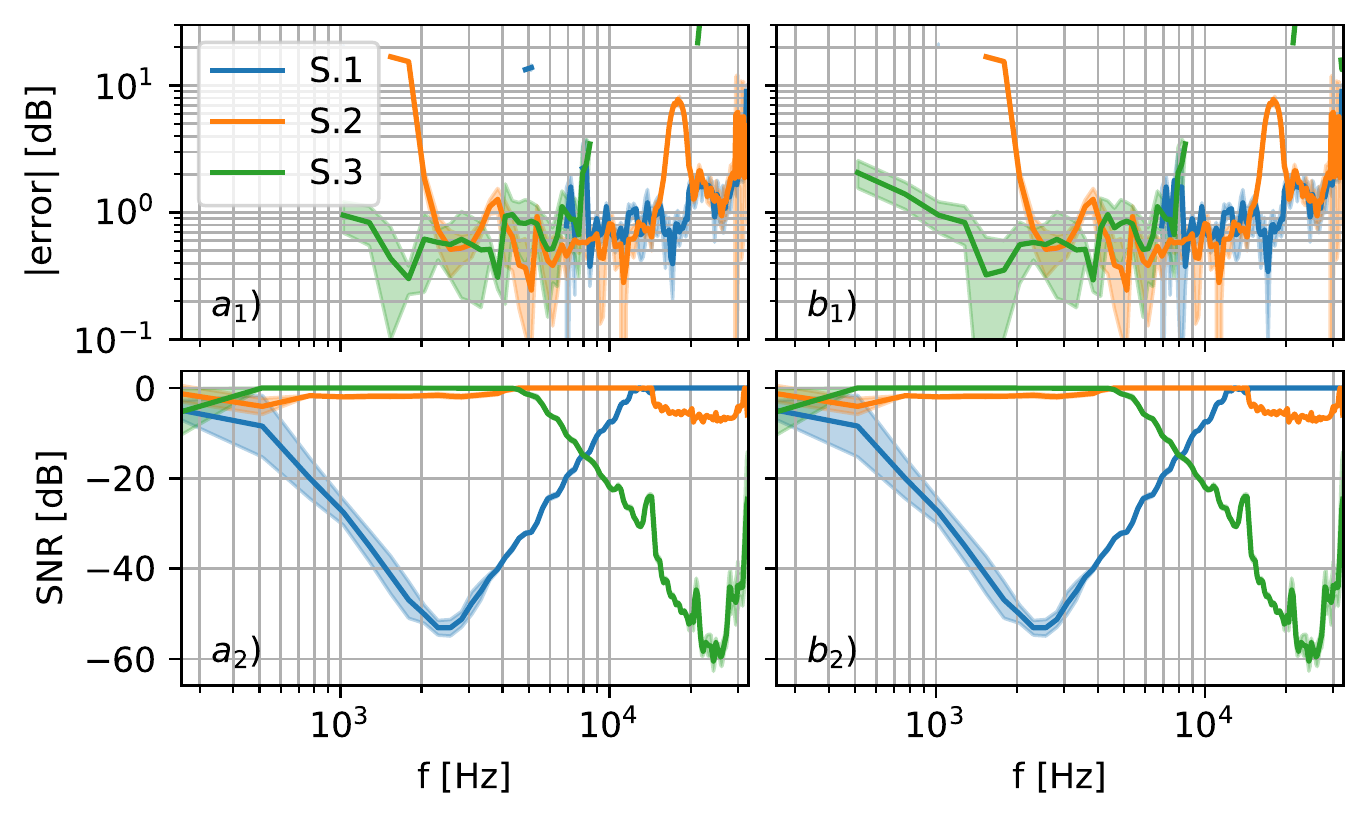}
	\caption{(Color online) The top row $(1)$ shows  the Mach-averaged absolute error $|\varepsilon|=|\langle\text{PSD}_\text{method}\rangle_M - \text{PSD}_\text{reference}|$ for each source in $a)$ for SIND and $b)$ for SIHC. The shaded area depicts the corresponding standard deviation. The bottom row $(2)$ shows the corresponding Signal To Noise ratios, see eq.~\ref{eq:SNR}.}
	\label{fig:Figure24}
\end{figure}

Third, a source-separation problem is created by superpositioning the three individual, denoised source CSMs for each Mach number and performing conventional beamforming in combination with CLEAN-SC on them, see Figure~\ref{fig:Figure23}. The individual sources are approximately $\Delta x_1=\SI{0.15}{\metre}$ apart. The performance of SIND and SIHC is evaluated on their ability to correctly detecting the dominant sources and by comparing the reconstructed spectra to the ground truth.\\

Both methods identify the three dominant monopole sources with the parameters given in table~\ref{tab:Table01}. Figure~\ref{fig:Figure24}, top row, shows the resulting absolute source power reconstruction error $|\varepsilon|$ for SIND and SIHC, and Table~\ref{tab:Table02} lists the frequency and Mach averaged reconstruction errors and averaged position errors $|\varphi|$. For the localization, both methods perform similarly on all sources with an estimation error that is smaller than two focus points. For the reconstruction of the corresponding spectra, both methods perform identically on source 2, similar on source 1 and different in terms of reconstructing the low frequencies on source 3, with SIHC performing slightly better.\\

\begin{figure}
	\centering
    \includegraphics[width=\reprintcolumnwidth]{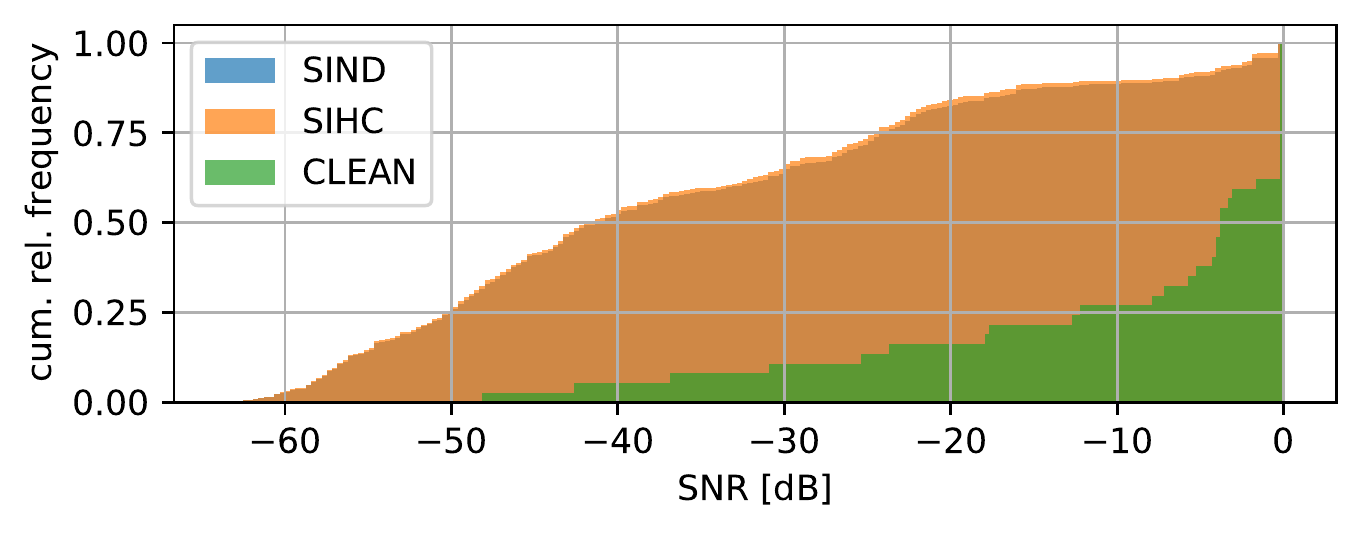}
	\caption{(Color online) The figure shows a cumulative histogram of the spectra SNR from all sources and Mach numbers, see eq.~\ref{eq:SNR}, that were not reconstructed $\text{SNR}(\text{PSD(f)}=\text{NaN})$.}
	\label{fig:Figure25}
\end{figure}

For evaluating the PSD reconstruction error in detail, two points will be considered. First, in real-world applications we often prefer a spectrum that is correct for high SNRs over small deviations at low SNRs. In this context, the SNR is the difference between the true single source's PSD and the summed PSD of all sources
\begin{equation}\label{eq:SNR}
    \text{SNR}_{S_i}(f,M) = \text{PSD}_{S_i} - \sum_{S_i} 10\log_{10} \left( 10^{\frac{\text{PSD}_{S_i}}{10}} \right) \,.
\end{equation}

Figure~\ref{fig:Figure24}, bottom row, shows the corresponding SNRs, see eq.~\ref{eq:SNR}. The SNR can also be interpreted as the per frequency normalized true source spectra from eq.~\ref{eq:ground_truth}. We can observe that at frequencies above $f\ge\SI{1}{\kilo\hertz}$ the spectra reconstruction failed or resulted in large errors when the SNR was low ($\text{SNR}\le-\SI{15}{\decibel}$). Since at high frequencies the beamforming map and resulting source-parts are normally well localized, see Figure~\ref{fig:Figure23}, and SIND and SIHC performed somewhat similar, see Figure~\ref{fig:Figure24}, we expect that these errors are mainly produced by the CLEAN-SC process itself. The relative interval of a valid spectrum reconstruction is important, that is for how many frequencies a solution is obtained, but not captured in the average error $|\varepsilon|$. The averaged relative frequency interval $0\le f_r\le 1$ is given in Table~\ref{tab:Table02}, a value of $f_r=1$ indicates that the spectra are reconstructed at each frequency $f$. Both methods perform similarly and reconstruct approximately $f_r=2/3$, while the individual CLEAN-SC references contain nearly full spectra. However, the CLEAN-SC reference was obtained from the individual source-maps (no CSM superposition) and thus the SNR was $\text{SNR}=\SI{0}{\decibel}$ (for high frequencies that were above the wind-tunnel noise-floor). Figure~\ref{fig:Figure25} shows the corresponding SNR of the parts of the source spectra, that were not reconstructed in a cumulative histogram. Thus, for each given SNR on the x-axis, the cumulative relative frequency shows how much percent of the failed reconstructions are below this SNR (e.g., \SI{50}{\percent} of the failed reconstructions are below $\text{SNR}\le-\SI{40}{\decibel}$ and $\SI{75}{\percent}$ are below $\text{SNR}\le-\SI{25}{\decibel}$). Both methods perform nearly identically on the generic dataset. The CLEAN-SC reference confirms, that the failed reconstructions are mainly due to the CLEAN-SC process.\\

\section{Discussion}
We presented two methods on how to detect sources and extract their spectra from sparse beamforming maps. The methods were developed and evaluated on real-world wind-tunnel datasets. The reason for this choice was that aeroacoustic experts only need support in identifying sources in beamforming maps of complex, ambiguous data. The drawback of this choice is the lack of a ground truth to quantify the results with a related metric. Thus, the results were only discussed qualitatively by comparing them to each other, their consistency, to the expectation of the aeroacoustic experts, and to the literature. Additionally, results of a generic dataset were analyzed quantitatively, which consisted of three superpositioned monopole point-sources with band-limited white noise with known source distribution and location and known emission power.\\

SIND was based on the idea that the source-parts' positions of compact acoustic sources at different frequencies appear spatially normal distributed in sparse beamforming maps. Thus, it yielded good results in finding point-like sources such as slat tracks, strakes, flap tracks, or the wingtip. SIND was also able to identify dense, overlapping sources like the flap side edge or point-like sources that were embedded in distributed sources such as the nacelle and the slat tracks in the inner slat region. It profited from stacking the histograms of multiple measurements at different Mach numbers and angles of attack to increase the sample size for the histogram, yet failed to recognize sparsely distributed source blobs with no clear midpoint. Wind-tunnel noise was a prominent example of this, as this source was projected on different parts of the image with increasing angle of attack $\alpha$ due to the mismatched focal plane. SINDs' results are robust against variations of the introduced thresholds and thus, were consistent with the expert out of the loop. The source positions on the two similar airframe models are consistent and based on the underlying source-part histogram we assume they are mostly correct. The correct identification of line-like sources, such as the slat, is ambiguous for this approach. In combination with CLEAN-SC, line-like sources' source-parts do not reassemble normal distributions. SIND tends to wrongly identify these as multiple point-like sources due to its distribution assumption in combination with CLEAN-SC processing. However, the airframe datasets showed that SIND's normal distribution approach was suited for most sources. For future improvements, a second distribution that is more suited towards fitting line-like sources is of interest. The use of DAMAS over CLEAN-SC might provide a more suited starting point for this. Also, SIND completely ignores the source-part's $\text{PSD}(f)$-information. Since the resulting spectra are expected to be smooth in a mathematical sense, this information could be potentially used additionally to the spatial criterion.\\

SIHC was based on the hierarchical clustering method HDBSCAN and thus did not assume a predefined source distribution. The source-parts were clustered directly in space, frequency, and SPL with the expert in the loop, as the results depend strongly on the set threshold. This means the correct threshold has to be determined manually to give accurate results. Because of the additional frequency and SPL information SIHC has the potential to separate spatially overlapping sources, such as slat tracks and slats. On the one hand, it clustered the inner slat and the flap side edge to single sources for which we assume the SIND solution to be more precise. On the other hand, it was able to identify sources containing source-parts that were too far scattered around the map to be identified by SIND, such as spurious noise sources that were not located on the wing. We consider this as an advantage, as these sources originated from the wind-tunnel and early in-situ detection during test measurements can potentially help to find and eliminate them.\\

Despite the similar identified source regions, SIND's estimation of individual source positions is more refined compared to the SIHC solution. While both methods identified the individual slat tracks (except for the A320 inner slat, where we assume the existence of two slat tracks, embedded in a distributed high-frequency noise source, see Figure~\ref{fig:Figure01}), the strakes and the wing tip on the Do728, SIHC missed the flap track closest to the wing tip on the Do728 and A320. It also clustered the inner slat region of the Do728 to a single ROI, as well as the nacelle region of the A320 and Do728 and the outer slat tip of the A320 and Do728.\\

The Do728 and A320 flap side edge, as well as an Do728 slat source, were shown in detail to evaluate the ROI quality. While the source-parts of the flap side edge form two overlapping normal distributions, SIHC identified a single source. We expect the flap side edge to be composed of multiple spatially distributed aeroacoustic source mechanisms~\citep{Howe1982,Dobrzynski2010} and showed that its spectrum is driven by at least two of them. Thus, we favor the SIND result over the SIHC result. The example Do728 slat source showed that the Strouhal number scaling tones are a distributed line source that is superimposed with point-like slat track sources which scale over Helmholtz number. While SIND identified most of the slat sources as point-like sources between the slat tracks, it was not able to assign the low-frequency source-parts to the slat that were located at the slat track positions. Since SIHC has the additional SPL and frequency information of each source-part and had no prior assumption of the source distribution it was able to assign the source-parts of overlapping sources to the correct sources in this case. Thus, we favor the SIHC result for the slat sources.\\

Both methods proved useful with different advantages and disadvantages to the real-world airframe datasets. SIHC works well for small datasets (e.g., a single angle of attack and few Mach variations) with little statistical noise. It is advantageous for exploring the dataset because a single threshold drastically changes the ROI outcome. Generally, density-based clustering methods tend to fail in separating clusters when too much noise is present that connects the clusters, so-called bridge points. Consequently, SIHC yields better results when decreasing the Welch block size, which increases the number of FFT averages and results in less statistical noise but also a lower frequency resolution. SIND works well for noisy datasets with high-resolution PSDs (large Welch block sizes) and yields stable results, that are mostly independent of the selected thresholds and profits from large datasets. Large datatsets ensure that each source is observed multiple times and thus, the total number of source-parts increases which allows the detection of sources that are not detectable in single noisy beamforming maps. Since SIND's thresholds only limit the processing time and drop sources after the identification, increasing or decreasing these values will not change the outcome of the remaining sources. Thus, SIHC is well-suited for an iterative process with the expert in the loop that can be fine-tuned to the desired outcome, while SIND requires no tuning to generate stable results and can be employed autonomously. The overall quality of SIND's results decreases with smaller datasets (fewer measurements), as the number of source-parts decreases, while SIHC's results improve, as it has to handle less statistical noise and vice versa. In specific cases, when two sources overlap spatially but can be distinguished based on their SPL$(f)$, such as slat sources, the SIHC method has a clear advantage over SIND, which naively assigns the source-parts based on their spatial probability alone. While dense source-distributions with bridge points are problematic for SIHC, it is able to detect sparse source-distributions without a clear midpoint, which SIND cannot detect (it relies on a well-localized distribution center as a starting point during the iterations). Thus, SIND's results heavily rely on a well-resolved beamforming map but can handle statistical noise due to insufficient CSM averages. SIHC, on the other hand, can to some degree correctly assign the source-parts that are far away from their corresponding source due to a low beamforming resolution based on their SPL and frequency information. However, its results suffer from statistical noise, so it requires long measurement times or small block sizes for sufficient CSM block averaging. It is possible to combine both methods by first employing SIND to extract the high-density clusters and then performing SIHC on the remaining source-parts.\\

The ability to recognize true sources, quantitatively estimate their position accuracy and acoustic power can only be evaluated on the generic dataset, where these quantities are known. The generic dataset provides a very limited source separation challenge as it consists only of spatially separated monopole sources. However, challenging aspects are the equidistant array spacing which results in strong grating lobes. These are even visible in the CLEAN-SC maps at high frequencies ($f\ge\SI{20}{\kilo\hertz}$), see Figure~\ref{fig:Figure23} between the true source positions. Also, the low array resolution (with a Rayleigh resolution limit $f_R\approx \SI{1}{\kilo\hertz}$ for the sources spaced around $\Delta x_1\approx\SI{0.15}{\metre}$) provides a separation challenge. These limitations resulted in CLEAN-SC failing to reconstruct the beamforming map at frequencies below $f_0\le\SI{1}{\kilo\hertz}$ or estimating the correct PSD at high frequencies ($\Delta \text{PSD}_{S_3}(f\approx\SI{30}{\kilo\hertz}) = -\SI{10}{\decibel}$) even when evaluating single source measurements as shown in Figure~\ref{fig:Figure23} and Figure~\ref{fig:Figure22}. For the assessment of the source localization and spectra reconstruction, only the combined error of CLEAN-SC and the proposed methods can be evaluated. However, since the CLEAN-SC maps of the individual sources are available, we spatially integrated these individual maps within a radius $r=\SI{0.1}{\metre}$ (reference area) around the true source locations, to obtain a CLEAN-SC reference position and spectrum of the sources. This allows an estimation of how much of the errors can be explained by the CLEAN-SC process, which is given in Table~\ref{tab:Table02}. Both proposed methods identified the three sources in the CLEAN-SC maps with similar spatial accuracy, see Table~\ref{tab:Table02}. The accuracy is overall higher than the average location of the source-parts within the CLEAN-SC reference area. Since the estimated source position is simply the average position of all assigned source-parts the position error is smaller for sources that have dominant high-frequency content than for sources that contain only low-frequency content (i.e. $S_3$). Out of the total spectrum range $f_r$ both proposed methods were able to reconstruct around $f_r\approx 2/3$ of the spectrum, see Table~\ref{tab:Table02}. Figure~\ref{fig:Figure24} and Figure~\ref{fig:Figure25} showed that most of these failed reconstructions happened at $S_1$ $f\le\SI{6}{\kilo\hertz}$ and $S_2$ $f\ge\SI{9}{\kilo\hertz}$, where the SNR is below $\text{SNR}\le-\SI{15}{\decibel}$. We expect these to be mainly caused not by the proposed methods confusing or missing source-parts but by the CLEAN-SC and conventional beamforming process on the superpositioned CSMs, which can be observed by the strong differences of the spectrum reconstructions in Figure~\ref{fig:Figure22} and Figure~\ref{fig:Figure24}, but also by the similarity of SIND and SIHC in Figure~\ref{fig:Figure24} and Figure~\ref{fig:Figure25}.\\

Performance-wise SIND is superior to SIHC and can be employed on datasets of any size. Additionally, both methods provide a confidence estimation for each source-part belonging to all sources. While the manual definition of ROIs simply determines if a source-part is part of a source or not, this information is valuable for an expert in estimating the reliability of the source spectra. Together, both methods cover the automatic source identification and spectrum generation from single, sparse low-resolution FFT beamforming maps to high-resolution FFT beamforming maps including multiple parameter variations with speed and accuracy that are unmatched by human experts. 
 
\section{Conclusion}
We presented the two methods ``SIND'' and ``SIHC'', which automatically detect aeroacoustic sources in deconvolved beamforming maps. They identify underlying source-distributions and thus, allow for the automatic determination of Regions Of Interest. To the best of our knowledge, these are the first automated approaches that can identify sources and generate corresponding spectra from sparse beamforming maps without prior information about the source locations. Both methods together cover a variety of real-world scenario used-cases, from single measurements with sparse source distributions to high-dimensional datasets with parameter variations and can be combined. Implementation details and results were discussed on scaled airframe half-model measurements and an error metric was introduced on a generic dataset featuring three known monopoles. In particular, the resulting Regions Of Interest and spectra of the flap side edge and a slat track were presented and showed that SIND is superior in separating dense, overlapping source regions, while SIHC is superior in assigning the source-parts to the correct sources which results in an improved reconstruction of spectra at low frequencies. For future work, SIND should be extended with a spectrum continuity criterion that ensures that the scattered low-frequency source-parts are assigned to the correct sources.

%\section*{Appendix}

\section*{Acknowledgments}
We thank the experts of the aeroacoustic group G\"ottingen, especially Dr. Thomas Ahlefeldt, for the helpful discussions on the analyzed beamforming results. We also acknowledge the DLR, Institute of Aerodynamics and Flow Technology, Department of Experimental Methods (contact: Carsten Spehr) for providing the SAGAS software which generated the beamforming and CLEAN-SC results for this paper. We thank the reviewers for their comments and insights which substantially improved this paper.

\bibliography{main}{}
\end{document}